\documentclass[%
 reprint,
 amsmath,amssymb,
 aps,
floatfix,
nobibline
]{revtex4-2}

\usepackage{graphicx}
\usepackage{dcolumn}
\usepackage{bm}
\usepackage{hyperref}

\begin{document}

\preprint{APS/123-QED}

\title{Wave propagation and scattering in time dependent media: Lippmann-Schwinger equations, multiple scattering theory, Kirchhoff Helmholtz integrals, Green's functions, reciprocity theorems and Huygens' principle}

\author{Xingguo Huang}
\author{Cong Wang}%
\author{Li Han}
 \email{Corresponding author:hanli23@jlu.edu.cn}
\affiliation{%
 College of Instrumentation and Electrical Engineering, Jilin University, Changchun, 130026, China
}%

\author{Stewart Greenhalgh}
\affiliation{
 Department of Earth Sciences, Institute of Geophysics, Swiss Federal Institute of Technology (ETH) Zürich, 8092 Zürich, Switzerland
}%
\author{Ru-Shan Wu}
\affiliation{%
 Modeling and Imaging Laboratory, Earth and Planetary Sciences, University of California, Santa Cruz, USA
}%


\date{\today}

\begin{abstract}
The use of waves to explore and analyze our environment is a common approach, spanning fields such as seismology, radar technology, biomedical imaging, and guided optics. In all these fields, wave scattering plays a central role for the modeling of complex wave propagation across all corners of science and engineering applications, including electromagnetics, acoustics, seismic and scattering physics. Wave control using time interfaces, where the properties of the medium through with the wave travels rapidly change in time, has opened further opportunities to control wave propagation in both space and time. For acoustic waves, studies on time modulated media have not been reported. In this context, full numerical solution of the wave equation using time interfaces is key to fully understand their potential. When applying time interfaces, the underlying physics of acoustic wave propagation and scattering and their similar roles on time and space, are still being explored. In this work, we introduce a mathematical formulation of the Lippmann-Schwinger integral equations for acoustic wave scattering when time interfaces are induced via a change of the velocity of the medium. We demonstrate that space-time duality for acoustic wave propagation with time interfaces and derive the Lippmann-Schwinger integral equations for wave scattering in time-dependent media, multiple scattering theory, Kirchhoff Helmholtz integrals, Green’s functions, reciprocity theorems. We experimentally verify our theoretical derivation by studying and measuring the acoustic wave scattering in strongly scattering media. We illustrate the proposed framework and present results of acoustic wave scattering without prior knowledge of the background wavefields. This improves the understanding of the generation and wave scattering and opens previously inaccessible research directions, potentially facilitating practical applications for acoustic, geophysical and optical imaging.
\end{abstract}

\maketitle


\section{\label{sec:level1}INTRODUCTION}

Understanding and controlling wave propagation has enabled the development of many areas including electromagnetism, acoustics, seismology and scattering physics, among others \cite{torrent2018loss,shaltout2019spatiotemporal,lyubarov2022amplified,sergeev2023ultrabroadband}, opening opportunities in new and improved approaches and applications including fully engineered wave manipulation, routing of information, resonant structures, photonics, filters and metamaterials, to name a few \cite{bruno2020negative,nassar2020nonreciprocity,li2021temporal,dikopoltsev2022light,engheta2023four,moussa2023observation}. A fundamental physical feature of wave propagation between two media is the existence, under certain conditions, of spatial reflections. Such reflections arise due to the spatial discontinuity between the materials with different parameters (impedance, velocity, density). Recently, emerging research has revealed the profound untapped potential of controlling wave propagation via time modulated media such as time interfaces. First proposed by \cite{morgenthaler1958velocity}, time interfaces (a rapid change of the properties of the material where a wave propagates) produce a time reflected (backward) and a time refracted (forward) waves, which are the temporal analogue of reflection and transmission when considering a spatial interface between two media \cite{engheta2023four,wang2023metasurface}]. Such time modulated media has become a hot research topic by its own merits with different groups recently reporting their applications from water to the optical regime \cite{morgenthaler1958velocity,fante1971transmission,bacot2015revisiting,caloz2019spacetime,mounaix2020time,galiffi2020wood,pacheco2020temporal,pacheco2021spatiotemporal,pacheco2021temporal,garnier2021wave,castaldi2021exploiting,ramaccia2021temporal,huidobro2021homogenization,pacheco2022time,li2022nonreciprocity,galiffi2022tapered,hidalgo2023damping,tirole2023double,rizza2023spin,kim2023unidirectional,moussa2023observation,pacheco2023time,pacheco2022time,pacheco2025temporal,wang2023controlling,pacheco2024spatiotemporal}. Interestingly, by exploiting two consecutive time interfaces applied at times close to each other (i.e., a square function with small time duration), instantaneous time mirrors (ITMs) have also been proposed and demonstrated for water waves \cite{bacot2016time,fink2017time}, revisiting the theoretical foundations of wave focusing and imaging. While traditional methods based on time reversal convert divergent waves into convergent waves that refocus at the source \cite{bal2019time}, ITMs produce a total backward wave (which is the combination of two waves produced by the two closely induced time interfaces) that refocus at the source. Although computationally tractable, a numerical algorithm to solve the nonlinear wave equation considering time interfaces is needed to further exploit such time modulated structures. The finite difference method has recently emerged as a leading candidate for solving parametric Partial Differential Equations, PDEs, addressing forward and inverse problems \cite{weglein1981scattering}. However, despite the immense promise of wave dynamics in the physics community, numerical solutions of time modulated media via time interfaces usually relies of commercial software that is not always accessible to everyone. 

 Here, we utilize the finite difference method for nonlinear problems to find solutions of the acoustic wave equation when time interfaces are introduced in the field of wave physics. We derive formulations of the Lippmann-Schwinger integral equations for acoustic wave scattering where time interfaces are induced. We consider that such time interfaces are applied by rapidly changing the velocity of the medium where the wave propagates in time. We introduce wave scattering, Green’s functions and time space duality. We derive the Lippmann Schwinger equations for multiple scattering theory, Kirchhoff Helmholtz integrals, Green’s functions, reciprocity theorems and Huygens’ principle. In particular, we present a full numerical solution of the wave equation using time interfaces and seismic records on the surface in strongly scattering media. Wave scattering problems are a class of forward and inverse problems in various fields \cite{wu1985multiple,wu1987diffraction,huang2023full}. Drawing motivation from nonlinear operators with neural networks, we recognize that the wavefields are differentiable with respect to time, this allows us to use automatic differentiation to formulate the second order derivatives. We will review the concept of time perturbation scattering problems. By rapidly changing the velocity of the medium in time, we are able to rapidly modify the wave time scattering properties at spatiotemporal interfaces. The results show interesting wave focusing and scattering of seismic wave with low frequency in the time domain. It is demonstrated that after time interfaces, two waves are created, one traveling forwards (diverging from the source) and one backwards (converging towards the source), and that the generated forward waves (FW) and backward waves (BW) have a new frequency due to the reflection and transmission of waves at the temporal discontinuity. The time-space seismic records show wave scattering and duality in the time domain.

\section{THEORY}
\subsection{Lippmann-Schwinger integral equation for wave scattering with spatial interfaces}
We start by briefly reviewing the Lippmann-Schwinger integral equation for wave scattering with spatial interfaces. Based on scattering theory, the scattered wavefields are the difference between the wavefields of the reference medium before the time interface and the wavefields of actual media during the time interface \cite{weglein1981scattering}. The Helmholtz equation is written as \cite{morse1946methods}
\begin{eqnarray}
\left ( \bigtriangledown ^{2} -\frac{1}{v^{2}\left ( \textbf{r} \right )} \frac{\partial^2 }{\partial t^2} \right )\psi =-S \left ( \textbf{r},t \right ),
\end{eqnarray}
where $\bigtriangledown$ displays the differentiation about the spatial position vector $\textbf{r}$, $S \left ( \textbf{r},t \right )$ represents the source function at an angular frequency $\omega$, $v\left ( \textbf{r} \right )$ is the wave propagation velocity.
The solution of the above equation $\psi \left ( \textbf{r} \right )$  is given by
\begin{eqnarray}
\psi(\mathbf{r},\omega)=-\int\mathcal{G}(\mathbf{r},\mathbf{r}^{\prime},\omega)S(\mathbf{r}^{\prime},\omega)d\mathbf{r}^{\prime}
\end{eqnarray}
where $\mathcal{G}(\mathbf{r},\mathbf{r}^{\prime})$ is the Green’s function at the location $\textbf{r}$ in the frequency domain. Defining $v_{0}(\mathbf{r})$ as the background medium velocity where the wave propagates, and the relationship of background wavefields $\psi^{0}(\mathbf{r})$ and scattered wavefields $\psi_s(\mathbf{r})$ is
\begin{eqnarray}
\psi_{s}(\mathbf{r})=\psi(\mathbf{r})-\psi^{(0)}(\mathbf{r}),
\end{eqnarray}
It can be concluded that \cite{morse1946methods}
\begin{eqnarray}
\left(\nabla^{2}-\frac{\omega^{2}}{\nu_{0}^{2}(\mathbf{r})}\right)G(\mathbf{r},\mathbf{r}^{\prime})=-\delta(\mathbf{r}-\mathbf{r}^{\prime})-O(\mathbf{r})G(\mathbf{r},\mathbf{r}^{\prime}),
\end{eqnarray}
with the scattering potential
\begin{eqnarray}
O(\mathbf{r},\omega)=\omega^{2}\left(\nu^{-2}(\mathbf{r})-\nu_{0}^{-2}(\mathbf{r})\right).	
\end{eqnarray}
The Lippmann-Schwinger integral equation is given by [\cite{morse1946methods,jakobsen2020convergent}]
\begin{equation}
\psi(\mathbf{r},\mathbf{r}^{\prime})=\psi^{(0)}(\mathbf{r},\mathbf{r}^{\prime})\\+\int d\mathbf{r}^{\prime\prime}G^{(0)}(\mathbf{r},\mathbf{r}^{\prime})V(\mathbf{r}^{\prime\prime})\psi(\mathbf{r}^{\prime\prime},\mathbf{r}^{\prime}),	
\end{equation}
where $G^{(0)}(\mathbf{r},\mathbf{r}^{\prime})$ is the background Green’s function, and it satisfies
\begin{eqnarray}
\left(\nabla^{2}-\frac{\omega^{2}}{v_{0}^{2}(\mathbf{r})}\right)\mathcal{G}^{(0)}(\mathbf{r},\mathbf{r}^{\prime})=-\delta(\mathbf{r}-\mathbf{r}^{\prime}),
\label{eq:seven}	
\end{eqnarray}
where $\delta$ is the Dirac delta point source function. Set Green’s function as the sum of the background Green’s function $\mathcal{G}^{(0)}(\mathbf{r},\mathbf{r}^{\prime})$ and the disturbance Green’s function $\Delta\mathcal{G}(\mathbf{r},\mathbf{r}^{\prime})$, and the $\Delta\mathcal{G}(\mathbf{r},\mathbf{r}^{\prime})$ satisfies
\begin{eqnarray}
\left(\nabla^{2}-\frac{\omega^{2}}{\nu_{0}^{2}(\mathbf{r})}\right)\Delta\mathcal{G}(\mathbf{r},\mathbf{r}^{\prime})=-O(\mathbf{r})\mathcal{G}(\mathbf{r},\mathbf{r}^{\prime}),
\label{eq:eight}
\end{eqnarray}
Eq.~(\ref{eq:seven}) and (\ref{eq:eight}) multiply $\Delta\mathcal{G}(\mathbf{r},\mathbf{r}^{\prime})$ and $G^{(0)}(\mathbf{r},\mathbf{r}^{\prime})$, respectively. Utilize the reciprocity theorem to exchange the positions of $\mathbf{r}^{\prime}$ and $\mathbf{r}$, and according to the relation $\Delta\mathcal{G}(\mathbf{r},\mathbf{r}^{\prime})=\mathcal{G}(\mathbf{r},\mathbf{r}^{\prime})-\mathcal{G}^{(0)}(\mathbf{r},\mathbf{r}^{\prime})$,
\begin{equation}
\begin{aligned}\mathcal{G}(\mathbf{r},\mathbf{r}^{\prime})&=\mathcal{G}^{(0)}\left(\mathbf{r},\mathbf{r}^{\prime}\right)+\int_{V}\mathcal{G}^{(0)}\left(\mathbf{r},\mathbf{r}^{\prime\prime}\right)\mathcal{O}(\mathbf{r}^{\prime\prime})\mathcal{G}(\mathbf{r}^{\prime\prime},\mathbf{r}^{\prime})\mathbf{d}\mathbf{r}^{\prime\prime}\\&+\int_{V}\left[\Delta\mathcal{G}(\mathbf{r}^{\prime\prime},\mathbf{r}^{\prime})\right]\nabla^{\prime2}\mathcal{G}^{(0)}(\mathbf{r},\mathbf{r}^{\prime\prime})\\&-\mathcal{G}^{(0)}(\mathbf{r},\mathbf{r}^{\prime\prime})\nabla^{\prime2}\left[\Delta\mathcal{G}(\mathbf{r}^{\prime\prime},\mathbf{r}^{\prime})\right]\mathrm{d}\mathbf{r}^{\prime\prime}\end{aligned}
\end{equation}
where $\int_{V}$ is the finite volume of three-dimensional space. Based on Green’s theorem,
\begin{equation}
\begin{aligned}\mathcal{G}(\mathbf{r},\mathbf{r}^{\prime})&=\mathcal{G}^{(0)}\left(\mathbf{r},\mathbf{r}^{\prime}\right)+\int_{V}\mathcal{G}^{(0)}\left(\mathbf{r},\mathbf{r}^{\prime\prime}\right)\mathcal{O}(\mathbf{r}^{\prime\prime})\mathcal{G}(\mathbf{r}^{\prime\prime},\mathbf{r}^{\prime})\mathrm{d}\mathbf{r}^{\prime\prime}\\&+\int_{S}\left[\Delta\mathcal{G}(\mathbf{r}^{\prime\prime},\mathbf{r}^{\prime})\right]\nabla^{\prime\prime}\mathcal{G}^{(0)}(\mathbf{r},\mathbf{r}^{\prime\prime})\\&-\mathcal{G}^{(0)}(\mathbf{r},\mathbf{r}^{\prime\prime})\nabla^{\prime\prime}\left[\Delta\mathcal{G}(\mathbf{r}^{\prime\prime},\mathbf{r}^{\prime})\right]\bullet n^{\prime\prime}\mathrm{d}\mathbf{r}^{\prime\prime}\end{aligned}
\end{equation}
where $n^{\prime\prime}$ is the outward normal unit vector of $S$ at $\mathbf{r}^{\prime\prime}$.

\subsection{Green's functions, reciprocity theorems}
In this section, we review the Green’s function and reciprocity theorems. The Green’s function is the solution of the Helmholtz equation
\begin{equation}
\bigtriangledown ^{2}\mathcal G -\frac{1}{c^{2}}\frac{\partial^2\mathcal G }{\partial t^2}=-4\pi \delta \left ( \mathbf{r} -\mathbf{r_{0}}\right )\sigma \left ( t-t_{0} \right )\\
\end{equation}
The initial conditions for this equation are $\mathcal G=0$ and $\frac{\partial \mathcal G  }{\partial x}=0$. The reciprocity relation can be written as
\begin{equation}
\bigtriangledown ^{2}\mathcal G -\frac{1}{c^{2}}\frac{\partial^2\mathcal G }{\partial t^2}=-4\pi \delta \left ( \mathbf{r} -\mathbf{r_{0}}\right )\sigma \left ( t-t_{0} \right )\\
\end{equation}
Considering the following equations
\begin{equation}
	\begin{aligned}
		\bigtriangledown ^{2}\mathcal G \left ( \textbf{r},t |\textbf{r}_{0} ,t_{0}\right ) -\frac{1}{c^{2}}\frac{\partial^2\mathcal G \left ( \textbf{r},t |\textbf{r}_{0} ,t_{0}\right ) }{\partial t^2}\\ =-4\pi \delta \left ( \mathbf{r} -\mathbf{r_{0}}\right )\sigma \left ( t-t_{0} \right )\\
	\end{aligned}
\end{equation}
and 
\begin{equation}
	\begin{aligned}
		\bigtriangledown ^{2}\mathcal G \left ( \textbf{r},-t |\textbf{r}_{1} ,-t_{1}\right ) -\frac{1}{c^{2}}\frac{\partial^2\mathcal G \left ( \textbf{r},-t |\textbf{r}_{0} ,-t_{1}\right ) }{\partial t^2}\\ =-4\pi \delta \left ( \mathbf{r} -\mathbf{r_{1}}\right )\sigma \left ( t-t_{1} \right )\\
	\end{aligned}
\end{equation}
Multiplying the first Equation by $\mathcal G \left ( \textbf{r},-t |\textbf{r}_{1} ,-t_{1}\right ) $ and multiplying the second equation by $\mathcal G \left ( \textbf{r},t |\textbf{r}_{0} ,t_{0}\right ) $ and integrating over time leads to 
\begin{equation}
	\begin{aligned}
		\int_{-\infty }^{t^{}}\int dV \mathcal G \left ( \textbf{r},t |\textbf{r}_{0} ,t_{0}\right ) \bigtriangledown ^{2}\mathcal G \left ( \textbf{r},-t |\textbf{r}_{1} ,-t_{1}\right )\\ - \mathcal G \left ( \textbf{r},-t |\textbf{r}_{1} ,-t_{1}\right )\bigtriangledown ^{2}\mathcal G \left ( \textbf{r},t |\textbf{r}_{0} ,t_{0}\right ) \\+\frac{1}{c^{2}}\mathcal G \left ( \textbf{r},t |\textbf{r}_{0} ,t_{0}\right ) \frac{\partial^2 }{\partial t^2}\mathcal G \left ( \textbf{r},-t |\textbf{r}_{1} ,-t_{1}\right )\\-\frac{1}{c^{2}}\mathcal G \left ( \textbf{r},-t |\textbf{r}_{1} ,-t_{1}\right ) \frac{\partial^2 }{\partial t^2}\mathcal G \left ( \textbf{r},t |\textbf{r}_{0} ,t_{0}\right )  \\
		= 4 \pi \left\{\mathcal G \left ( \textbf{r}_{0},-t_{0} |\textbf{r}_{1} ,-t_{1}\right )-\mathcal G \left ( \textbf{r}_{1},t_{1} |\textbf{r}_{0} ,t_{0}\right ) \right\}
	\end{aligned}
\end{equation}

Applying the Green's theorem to the left hand of the equation, we get 
\begin{equation}
	\begin{aligned}
		\frac{\partial }{\partial t}\left [\mathcal G \left ( \textbf{r},t |\textbf{r}_{0} ,t_{0}\right )  \frac{\partial }{\partial t}\mathcal G \left ( \textbf{r},-t |\textbf{r}_{1} ,-t_{1}\right ) \right. \\ 
		\left. -\mathcal G \left ( \textbf{r},-t |\textbf{r}_{1} ,-t_{1}\right ) \frac{\partial }{\partial t} \mathcal G \left ( \textbf{r},t |\textbf{r}_{0} ,t_{0}\right ) \right ] \\  
		=\frac{\partial }{\partial t}\left [\mathcal G \left ( \textbf{r},t |\textbf{r}_{0} ,t_{0}\right )  \frac{\partial^{2} }{\partial t^{2}}\mathcal G \left ( \textbf{r},-t |\textbf{r}_{1} ,-t_{1}\right ) \right. \\
		\left. -\mathcal G \left ( \textbf{r},-t |\textbf{r}_{1} ,-t_{1}\right ) \frac{\partial^2 }{\partial t^2 } \mathcal G \left ( \textbf{r},t |\textbf{r}_{0} ,t_{0}\right ) \right ]
	\end{aligned}
\end{equation}
Then we have the following equation for the left part 
\begin{equation}
	\begin{aligned}
		\int_{-\infty }^{t^{}}\int d\textbf{S} \mathcal G \left ( \textbf{r},t |\textbf{r}_{0} ,t_{0}\right ) \bigtriangledown ^{2}\textrm{grad} \mathcal G \left ( \textbf{r},-t |\textbf{r}_{1} ,-t_{1}\right ) \\
		 - \mathcal G \left ( \textbf{r},-t |\textbf{r}_{1} ,-t_{1}\right )\bigtriangledown ^{2}\mathcal G \left ( \textbf{r},t |\textbf{r}_{0} ,t_{0}\right ) \\
		+\frac{1}{c^{2}}\int dV\left [ \mathcal G ( \textbf{r},t |\textbf{r}_{0} ,t_{0}\right ) \frac{\partial }{\partial t}\mathcal G \left ( \textbf{r},-t |\textbf{r}_{1} ,-t_{1}\right )\\-\frac{1}{c^{2}}\mathcal G \left ( \textbf{r},-t |\textbf{r}_{1} ,-t_{1}\right ) \frac{\partial }{\partial t}\mathcal G \left ( \textbf{r},t |\textbf{r}_{0} ,t_{0}\right ) ]  \\
	\end{aligned}
\end{equation}

The Green's functions is the solution of the equation
\begin{equation}
\nabla_0^2 \psi\left(\mathrm{r}_{\mathrm{v}}, t_0\right)-\frac{1}{c^2} \frac{\partial^2 \psi}{\partial t_0^2}=-4 \pi q\left(\mathrm{r}_0, t_0\right)
\end{equation}
and
\begin{equation}
	\begin{aligned}
		\nabla_0^2  \mathcal G \left(\mathbf{r}, t \mid \mathbf{r}_0, t_0\right)-\frac{1}{c^2} \frac{\partial^2  \mathcal G \left(\mathbf{r}, t \mid \mathbf{r}_0, t_0\right)}{\partial t_0^2}\\ =-4 \pi \delta\left(\mathbf{r}-\mathbf{r}_0\right) \delta\left(t-t_0\right)
	\end{aligned}
\end{equation}
Then we have 
\begin{equation}
	\begin{aligned}
\int_0^{t^{+}} d t_0 \int d V_0\left\{ \mathcal G \nabla_0^2 \psi-\psi \nabla_0^2  \mathcal G+\right. \left.\frac{1}{c^2}\left(\frac{\partial^2  \mathcal G }{\partial t_0^2} \psi- \mathcal G \frac{\partial^2 \psi}{\partial t_0^2}\right)\right\} \\ =4 \pi\left\{\psi(\mathbf{r}, t)-\int_0^{t^{+}} d t_0 \int d V_0 q\left(\mathbf{r}_0, t_0\right)  \mathcal G \right\}
	\end{aligned}
\label{eq:twenty}
\end{equation}
Applying Green's theorem to Eq.~(\ref{eq:twenty}), we obtain
\begin{equation}
	\begin{split}
		4 \pi \psi(\mathbf{r}, t) &= \int_0^{t^{+}} \mathrm{d} t_0 \oint \mathrm{dS}_0 \cdot \left(G \operatorname{grad}_0 \psi - \psi \operatorname{grad}_0 \mathcal{G} \right) \\
		&\quad + \frac{1}{c^2} \int \mathrm{d} V_0 \left[\frac{\partial \mathcal{G}}{\partial t_0} \psi - \mathcal{G} \frac{\partial \psi}{\partial t_0}\right]_0^{t^{+}} \\
		&\quad + 4 \pi \int_0^{t^{+}} \mathrm{d} t_0 \int \mathrm{d} V_0 q\left(\mathbf{r}_0, t_0\right) \mathcal{G}
	\end{split}
\end{equation}
It follows that 
\begin{equation}
	\begin{aligned}
		4 \pi \psi(\mathbf{r}, t) &= 4 \pi \int_0^{t^{+}} \mathrm{d} t_0 \int V_d \, \mathrm{d}\left(\mathbf{r}, t \mid \mathbf{r}_0, t_0\right) q\left(\mathbf{r}_0, t_0\right) \\
		&\quad + \int_0^{t^{+}} \mathrm{d} t_0 \oint \mathrm{dS}_0 \cdot \left( \mathcal{G} \operatorname{grad}_0 \psi - \psi \operatorname{grad}_0 \mathcal{G} \right) \\
		&\quad - \frac{1}{c^2} \int \mathrm{d} V_0 \left[ \left( \frac{\partial \mathcal{G}}{\partial t_0} \right)_{t_0=0} \psi_0\left(\mathbf{r}_0\right) - \mathcal{G}_{t_0 \sim 0} v_0\left(\mathbf{r}_0\right) \right]
	\end{aligned}
\end{equation}
\subsection{Huygen's Principle}
In the 19th century, Kirchhoff, Helmholtz and Rayleigh derived mathematical expressions which formalize Huygens' principle. The  Huygens' principle is widely used in wave propagation in geophysics. The basic idea is that all points in the opening surface act as secondary sources  and the superposition of the waves converge to the wave radiated
by the original source. The integral in the equation above can be written as 
\begin{equation}
	\begin{aligned}
		\psi(\mathbf{r}, t) &= \frac{1}{4 \pi} \int_0^{t^{+}} \mathrm{d} t_0 \oint \mathrm{d} \mathbf{S}_0 \cdot \bigg\{ 
		\left[\frac{\delta\left(t_0 - t + R / c\right)}{R}\right] \operatorname{grad}_0 \psi \\
		&\quad - \psi \operatorname{grad}_0 \left[\frac{\delta\left(t_0 - t + R / c\right)}{R}\right] \bigg\}
	\end{aligned}
\end{equation}
Integrating over $t_0$ leads to 
\begin{equation}
	\begin{array}{ccl}
		\begin{aligned}
		\int_0^{t^{+}}\left(\frac{1}{R}\right) \delta\left(t_0-t+\frac{R}{c}\right) \operatorname{grad}_0 \psi\left(\mathrm{r}_0, t_0\right) d t_0\\=\left(\frac{1}{R}\right)\operatorname{grad}_0 \psi\left(\mathrm{r}_0, t-\frac{R}{c}\right)
		\end{aligned}
	\end{array}
\end{equation}
Then we have 

\begin{equation}
\begin{split}
	\int_0^{t^{+}} \psi \, &\operatorname{grad}_0\left[\frac{\delta(t_0-t+R/c)}{R}\right] dt_0 \\
	&= \int_0^{t^{+}} \psi \frac{\partial}{\partial R}\left[\frac{\delta(t_0-t+R/c)}{R}\right] \operatorname{grad}_0 R \, dt_0 \\
	&= \int_0^{t^{+}} \psi(\mathbf{r}_0, t_0) \left(\frac{{R}}{R^3}\right) \left[-\delta\left(t_0-t+\frac{R}{c}\right) \right. \\ 
	&\quad \left. + \frac{R}{c} \delta'\left(t_0-t+\frac{R}{c}\right)\right] dt_0 \\
	&= -\frac{{R}}{R^3} \left\{ \psi\left(\mathbf{r}_0, t-\frac{R}{c}\right) + \frac{R}{c} \left[ \frac{\partial}{\partial t_0} \psi(\mathbf{r}_0, t_0) \right]_{t_0 = t - R/c} \right\}
\end{split}	
\end{equation}
The wavefields can be written as a surface integral 
\begin{equation}
	\begin{split}
		\psi(\mathbf{r}, t) = \frac{1}{4 \pi} \oint & \mathrm{d}\mathbf{S}_0 \cdot \bigg[
		\left(\frac{1}{R}\right) \operatorname{grad}_0 \psi\left(\mathbf{r}_0, t_0\right) \\
		& + \left(\frac{\mathbf{R}}{R^3}\right) \psi\left(\mathbf{r}_0, t_0\right) \\
		& - \left(\frac{R}{c R^2}\right) \frac{\partial}{\partial t_0} \psi\left(\mathbf{r}_0, t_0\right) \bigg]_{t_0 = t - (R/c)}
	\end{split}
\end{equation}

\subsection{Kirchhoff Helmholtz integrals}
The standard acoustic Kirchhoff integral can be written as
\begin{equation}
	\begin{split}
		\check{U}(G, \omega) = \frac{1}{4 \pi} \iint_{\Sigma} \mathrm{d}\Sigma \, \frac{1}{\varrho} \bigg[ 
		&\check{\mathcal{G}}(G, P, \omega) \frac{\partial \check{U}}{\partial n}(P, \omega) \\
		&- \check{U}(P, \omega) \frac{\partial \check{\mathcal{G}}}{\partial n}(G, P, \omega) \bigg]
	\end{split}
\end{equation}
where the vector $\hat{\boldsymbol{n}}$, normal to $\Sigma$, points outwards, i.e., out of the enclosing surface $\Sigma$ into the region where the sources of the wavefield $\check{U}(G, \omega)$ are found. Also, $\partial / \partial n=\hat{\boldsymbol{n}} \cdot \hat{\boldsymbol{\nabla}}$ denotes the normal derivative in that direction. Finally, $\varrho$ is the density of the medium at point $P$. The integral for transmitted waves can be written as

\begin{equation}
	\begin{split}
		\check{U}^t\left(\hat{\boldsymbol{r}}_G, \omega\right) = \frac{-1}{4 \pi} \int_{\Sigma_T} & \mathrm{d}\Sigma_T \, f_m(\hat{\boldsymbol{r}}) \bigg[ \\
		& \check{\mathcal{G}}\left(\hat{\boldsymbol{r}}, \omega ; \hat{\boldsymbol{r}}_G\right) \frac{\partial \check{U}^t(\hat{\boldsymbol{r}}, \omega)}{\partial n} \\
		& - \check{U}^t(\hat{\boldsymbol{r}}, \omega) \frac{\partial \check{\mathcal{G}}\left(\hat{\boldsymbol{r}}, \omega ; \hat{\boldsymbol{r}}_G\right)}{\partial n} \bigg]
	\end{split}
\end{equation}
where $\breve{U}^t(\hat{r}, \omega)$ inside the integral represents the wavefield at the transmitting interface $\Sigma_T$ directly after transmission.
\begin{equation}
	\begin{split}
		\check{U}^r\left(\hat{\boldsymbol{r}}_G, \omega\right) = \frac{-1}{4 \pi} \int_{\Sigma_R} & \mathrm{d}\Sigma_R \, f_m(\hat{\boldsymbol{r}}) \bigg[ \\
		& \check{\mathcal{G}}\left(\hat{\boldsymbol{r}}, \omega ; \hat{\boldsymbol{r}}_G\right) \frac{\partial \check{U}^r(\hat{\boldsymbol{r}}, \omega)}{\partial n} \\
		& - \check{U}^r(\hat{\boldsymbol{r}}, \omega) \frac{\partial \check{\mathcal{G}}\left(\hat{\boldsymbol{r}}, \omega ; \hat{\boldsymbol{r}}_G\right)}{\partial n} \bigg]
	\end{split}
\end{equation}

Note that the Kirchhoff integral equation for reflected waves does not depend on whether the surface $\Sigma_R$ of $R$ is a closed or an open surface.

Defining $ \Delta  \mathcal G \left(\mathbf{r}, \mathbf{r}^{\prime}\right)= \mathcal G \left(\mathbf{r}, \mathbf{r}^{\prime}\right)- \mathcal G_{0}\left(\mathbf{r}, \mathbf{r}^{\prime}\right)$, then we have
\begin{equation}
	\begin{aligned}
		\mathcal{G}\left(\mathbf{r}, \mathbf{r}^{\prime}\right) &= \mathcal{G}\left(\boldsymbol{r}, \boldsymbol{r}^{\prime}\right) - \mathcal{G}_{0}\left(\boldsymbol{r}, \boldsymbol{r}^{\prime}\right) \\
		&= \int_{x, y \in R_{\mathrm{m}}} \bigg[ \mathcal{G}\left(x^{\prime \prime}, z^{\prime \prime}, \boldsymbol{r}^{\prime}\right) \nabla^{\prime \prime 2} \mathcal{G}_{0}\left(\boldsymbol{r}, x^{\prime \prime}, z^{\prime \prime}\right) \\
		&\quad - \mathcal{G}_{0}\left(\boldsymbol{r}, x^{\prime \prime}, z^{\prime \prime}\right) \nabla^{\prime \prime 2} \mathcal{G}\left(x^{\prime \prime}, z^{\prime \prime}, \boldsymbol{r}^{\prime}\right) \bigg] \mathrm{d} x^{\prime \prime}
	\end{aligned}
\end{equation}
or
\begin{equation}
	\begin{aligned}
		\Delta \mathcal{G}\left(\boldsymbol{r}, \boldsymbol{r}^{\prime}\right) 
		&= \mathcal{G}\left(\boldsymbol{r}, \boldsymbol{r}^{\prime}\right) - \mathcal{G}_{0}\left(\boldsymbol{r}, \boldsymbol{r}^{\prime}\right) \\
		&= \int_{S_{\mathrm{R}}} \bigg[ 
		G\left(x^{\prime \prime}, z^{\prime \prime}, \boldsymbol{r}^{\prime}\right) \nabla^{\prime \prime} \mathcal{G}_{0}\left(\boldsymbol{r}, x^{\prime \prime}, z^{\prime \prime}\right) \\
		&\quad - \mathcal{G}_{0}\left(\boldsymbol{r}, x^{\prime \prime}, z^{\prime \prime}\right) \nabla^{\prime \prime} \mathcal{G}\left(x^{\prime \prime}, z^{\prime \prime}, \boldsymbol{r}^{\prime}\right) 
		\bigg] \cdot n^{\prime \prime} \, \mathrm{d} x^{\prime \prime}
	\end{aligned}
\end{equation}
It can be written as
\begin{equation}
	\begin{aligned}
		\Delta \mathcal{G}\left(\boldsymbol{r}, \boldsymbol{r}^{\prime}\right) 
		&= \mathcal{G}\left(\boldsymbol{r}, \boldsymbol{r}^{\prime}\right) - \mathcal{G}_0\left(\boldsymbol{r}, \boldsymbol{r}^{\prime}\right) \\
		&= \int_{S_{-m}} \Bigl[ \mathcal{G}\left(x^n, z^n, \boldsymbol{r}^{\prime}\right) \nabla^n \mathcal{G}_0\left(\boldsymbol{r}, x^n, z^n\right) \\
		&\qquad - \mathcal{G}_0\left(\boldsymbol{r}, x^n, z^n\right) \nabla^n \mathcal{G}\left(x^n, z^n, \boldsymbol{r}^{\prime}\right) \Bigr] \cdot \boldsymbol{n}^n \, \mathrm{d} x^n \\
		&= \int_{z=z_n} \Bigl[ \mathcal{G}\left(x^n, z^n, \boldsymbol{r}^{\prime}\right) \nabla^n \mathcal{G}_0\left(\boldsymbol{r}, x^n, z^n\right) \\
		&\qquad - \mathcal{G}_0\left(\boldsymbol{r}, x^n, z^n\right) \nabla^n \mathcal{G}\left(x^n, z^n, \boldsymbol{r}^{\prime}\right) \Bigr] \cdot \boldsymbol{n}^n \, \mathrm{d} x^n
	\end{aligned}
\end{equation} 

if$ S_{z_{m}+\Delta z} \rightarrow \infty $, then we have the Kirchhoff Helmholtz integral
\begin{equation}
	\begin{aligned}
		\Delta \mathcal{G}(\boldsymbol{r}, \boldsymbol{r}') &= \mathcal{G}(\boldsymbol{r}, \boldsymbol{r}') - \mathcal{G}_0(\boldsymbol{r}, \boldsymbol{r}') \\
		&= \int_{S_{-m}} \Bigl[ \mathcal{G}(x^n, z^n, \boldsymbol{r}') \nabla^n \mathcal{G}_0(\boldsymbol{r}, x^n, z^n) \\
		&\quad - \mathcal{G}_0(\boldsymbol{r}, x^n, z^n) \nabla^n \mathcal{G}(x^n, z^n, \boldsymbol{r}') \Bigr] \cdot \boldsymbol{n}^n \, \mathrm{d} x^n \\
		&= \int_{z=z_n} \Bigl[ \mathcal{G}(x^n, z^n, \boldsymbol{r}') \nabla^n \mathcal{G}_0(\boldsymbol{r}, x^n, z^n) \\
		&\quad - \mathcal{G}_0(\boldsymbol{r}, x^n, z^n) \nabla^n \mathcal{G}(x^n, z^n, \boldsymbol{r}') \Bigr] \cdot \boldsymbol{n}^n \, \mathrm{d} x^n
	\end{aligned}
\end{equation}

\subsection{Lippmann-Schwinger integral equation for wave scattering with time interfaces}
\begin{figure}
	\centering
	\includegraphics[scale=0.45]{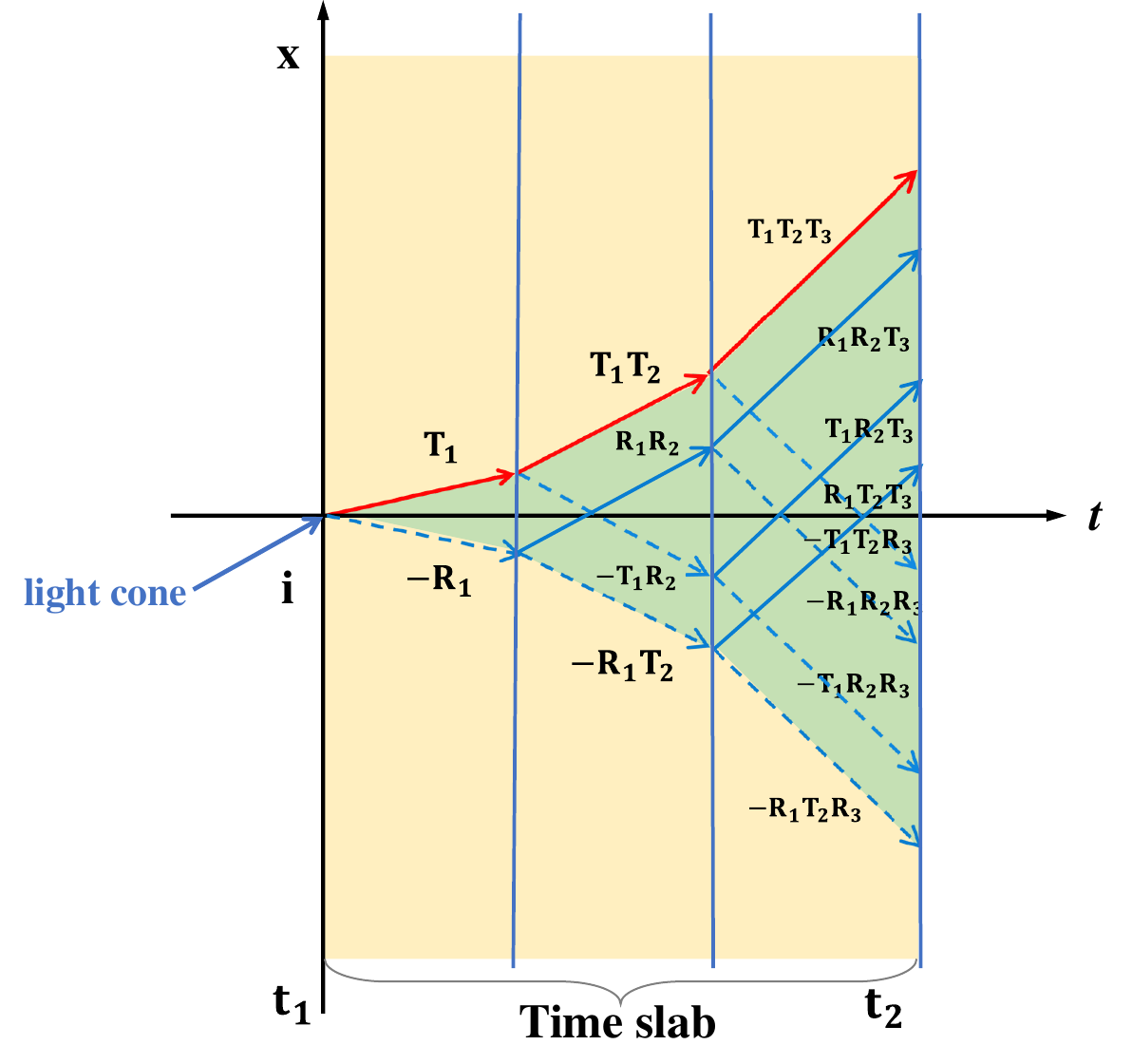}
	\caption{\label{fig:1} Schematics of wave scattering on temporal slabs.}
\end{figure}

Here, we consider wave scattering in the time varying media where time interfaces are induced (Fig.\ref{fig:1}), and present arithmetic solutions for wavefields in such media in the next section. As explained in the Introduction, we consider that the velocity of the entire medium where the wave propagates varies in time following a rapid modulation from one value to a different value at times   with  , i.e., different time interfaces. In such a time varying medium, there is a new source term about the incoming wavefields at times. To account for this, we consider the following time-perturbation wave equation in time varying media [\cite{bacot2016time}]
\begin{equation}
\nabla^{2}-\frac{1}{\nu^{2}}\frac{\partial^{2}}{\partial t^{2}}u(\mathbf{r},t)=-\frac{\alpha}{\nu^{2}}\delta(t-t_{m})\frac{\partial^{2}u(\mathbf{r},t)}{\partial t^{2}}
\label{eq:thirty-four}
\end{equation}	
where the new source term $f(\mathrm{r},t)=-\frac{\alpha}{\nu_{0}^{2}}\delta(t-t_{m})\frac{\partial^{2}u(\mathrm{r},t)}{\partial t^{2}}$, and the value of $\alpha$ is larger that the perturbation to the velocity of the medium is smaller. The source provides a singularity that acts as a delta function at a specific moment in time. While the source term is spatially distributed, it remains temporally localized. Consequently, a signal generated at a given time and perturbed at a previous time—corresponding to one period before the time interface—can refocus at its origin if the velocity remains unchanged after the time interface. Mathematically, the time derivative of the original signal is satisfied by the acoustic wave equation where time interfaces are present. This is because the right-hand side of Eq.~(\ref{eq:thirty-four}) shows a velocity perturbation in time  . Importantly, the initial state of waves is known to decompose into a combination of the unperturbed wavefields and an additional set of perturbed wavefields [\cite{bacot2016time,fink2017time}].

Wave reflection on the space slab and time slab is illustrated in Fig.\ref{fig:1}, where it is shown how a spatial slab produces a total transmitted and reflected signal which is the result of multiple reflections occurring within the slab. For the time slab, however, four signals are produced after applying the two time interfaces: two traveling forward and two traveling backward (i.e., multiple reflections do not exist in the temporal case). The goal of the forward scattering problem is to develop numerical solutions for a time-dependent perturbation scattering theory. To address this, we derive scattering integral equations for wave scattering in time varying media based on the time-dependent wave equation. Assuming  $V=\frac{1}{v^{2}}-\frac{1}{v_{0}^{2}}$(with $v_{0}$ as the background velocity and $v$ as the actual velocity.) and inserting into Eq.~(\ref{eq:thirty-four}), we obtain
\begin{multline}
	\left(\nabla^{2}-\frac{1}{\nu_{2}^{0}}\frac{\partial^{2}}{\partial t^{2}}-V(\mathbf{r})\frac{\partial^{2}}{\partial t^{2}}\right. \\
	\left. +\frac{\alpha}{\nu_{2}^{0}}(t-t_{m})\frac{\partial^{2}u(\mathbf{r},t)}{\partial t^{2}}\right) u(\mathbf{r},t) = s(t).
	\label{eq:thirty-five}
\end{multline}
Then, we set the operators
\begin{equation}
	L=\left(\nabla^{2}-\frac{1}{v^{2}}\frac{\partial^{2}}{\partial t^{2}}\right),
\end{equation}
and
\begin{equation}
	L_{0}=\left(\nabla^{2}-\frac{1}{v_{2}^{0}}\frac{\partial^{2}}{\partial t^{2}}\right),
\end{equation}

the scattering potential is
\begin{equation}
	L-L_{0}=V\frac{\partial^{2}}{\partial t^{2}}+\frac{\alpha}{\nu_{0}^{2}}\delta(t-t_{m})\frac{\partial^{2}u(\mathrm{x},t)}{\partial t^{2}}
\end{equation}
The scattering solution refers to the Lippmman-Schwinger equation where the time interfaces are present, which is written as
\begin{equation}
	\begin{split}
		u(\mathbf{x}) &= u^{(0)}(\mathbf{r}) + \int\limits_{\Omega} \mathrm{d}\mathbf{x}^{\prime} \, G^{(0)}(\mathbf{x} - \mathbf{x}^{\prime}) \bigg( V\frac{\partial^{2}}{\partial t^{2}} \\
		&\qquad + \frac{\alpha}{v_{0}^{2}}\delta(t-t_{m})\frac{\partial^{2}u(\mathbf{r},t)}{\partial t^{2}} \bigg) u(\mathbf{r}^{\prime})
	\end{split}
	\label{eq:thirty_nine}
\end{equation}

The instantaneous source, which is proportional to the second order time derivative of the wavefields, provides a singularity which represents a delta function at a given time. The source term is delocalized in space but localized in time. The concept of the time varying media is not new, and time reflection/refraction waves in the time varying media can be obtained by giving a strong perturbation at a given time the medium. Indeed, a signal produced at time $t$ = 0 and perturbed at $t$ = T, can refocus at its source at time $t$ = 2T[\cite{bacot2016time,fink2017time}]. The time derivative of the original signal associated with refocusing waves satisfied the time perturbation wave equation. Importantly, the new initial state of waves is known to decompose into a superposition of the unperturbed wavefields as well as an additional perturbed wavefields. To utilize a numerical solution to model Eq.~(\ref{eq:thirty_nine}), it is significant to have a basic understanding of wave scattering.
\subsection{Wave reflection and refraction at time and space interfaces and Green’s function}
Next, we discuss wave reflection and refraction at time and space interfaces in the context of plane wave propagation. The velocity is changed from $v_{1}$ to $v_{2}$ at $t=t_{1}$, and there is a time reflection (backward wave) and time refraction (forward wave) of wavefield $u$. Note that there are two waves at the new frequency $\frac{v_{2}}{v_{1}}\omega$. The two waves have the same wavenumber $\frac{\omega}{v_{1}}$[\cite{morgenthaler1958velocity}]. When the wave hits a spatial interface at $x=0$, there is another wave that propagates in the direction $k_{1}^{\prime}=x\cos\theta_{1}+z\sin\theta_{1}$. The incident wavefields can be written as [\cite{fante1971transmission}]
\begin{equation}
	u=p_{0}exp\left[i\left(\omega t-k_{0}\cdot r\right)\right]u\left(t-\frac{\hat{k}_{0}\cdot r}{\nu}\right)
\end{equation}
where $p_{0}$ is a initial wavefield, $\omega$ is the angular frequency, $k_{0}$ is initial wave vector, $t$ is the propagation time. Then the formula for the fields for times $0<t<t_{1}$ is written as
\begin{equation}
	u_{1}=p_{1}T_{1}exp\left[i\left(\omega t-k_{1}\cdot r\right)\right]u\left(t-\frac{\hat{k}_{1}\cdot r}{\nu_{1}}\right)
\end{equation}
where $p_{}$ is wavefield, $p_{1}=z\cos\theta_{1}+x\sin\theta_{1}$, and $k_{1}$ is incident wave vector.

Different from spatial interfaces, the reflection and refraction waves both change after time interfaces. The refraction plane wave is given by [\cite{wapenaar2025green}]
\begin{equation}
	u_{\mathrm{FW}}=F_{u}T_{2}exp\left[i\left(\omega t-k_{2}\cdot r\right)\right]u\left(t-\frac{\hat{k}_{2}\cdot r}{\nu}\right)
\end{equation}
\begin{equation}
	u_{\mathrm{1FW}}=F_{u}T_{2}exp\left[i\left(\omega t-k_{2}\cdot r\right)\right]u\left(t-\frac{\hat{k}_{2}\cdot r}{\nu_{2}}\right)
\end{equation}
where $F_{_{\mathrm{u}}}$ is the refraction coefficient for the wavefield $u$ , $k_{2}=x\cos\theta_{2}-z\sin\theta_{2}$. The reflection wave for $t>t_1$ is expressed as
\begin{equation}
	u_{\mathrm{BW}}=B_{u}exp\left[i\left(\omega t+k_{2}\cdot r\right)\right]u\left(t-\frac{\hat{k}_{2}\cdot r}{\nu}\right)
\end{equation}
\begin{equation}
	u_{\mathrm{lBW}}=B_{u}T_{2}exp\left[i\left(\omega t+k_{2}\cdot r\right)\right]u\left(t-\frac{\hat{k}_{2}\cdot r}{\nu_{2}}\right)
\end{equation}
where $B_{u}$ is the reflection coefficient for the wavefield $u$.
Then the formula for the split plane waves in time can be written as [\cite{fante1971transmission}]
\begin{widetext}
\begin{equation}
	\begin{aligned}
			\mathrm{u_{2}}&=p_{1}A_{1}\left\{u\left(t-t_{1}+\frac{\nu_{1}}{\nu_{2}}t_{1}-\frac{k_{1}\cdot r}{\nu_{2}}\right)-u\left(t-t_{1}-\frac{\hat{k}_{1}\cdot r}{\nu_{2}}\right)\right\}\exp\left[i\frac{\nu_{2}}{\nu_{1}}\omega\left(t-t_{1}\right)\right]\exp\left(i\omega t_{1}\right)\cdot\exp\left(-ik_{1}\cdot r\right)\\&+p_{1}A_{2}\left\{u\left(t-t_{1}+\frac{\hat{k}_{1}\cdot r}{\nu_{2}}\right)-u\left(t-t_{1}-\frac{\nu_{1}}{\nu_{2}}t_{1}+\frac{\hat{k}_{1}\cdot r}{\nu_{2}}\right)\right\}exp\left[-i\frac{\nu_{2}}{\nu_{1}}\omega(t-t_{1})\right]\cdot exp\left(i\omega t_{1}\right)exp\left(-ik_{1}\cdot r\right)\\&+q_{1}A_{3}\left\{u\left(t-t_{1}-\frac{\hat{k}_{1}^{^{\prime}}\cdot r}{\nu_{2}}\right)-u\left(t-t_{1}-\frac{\nu_{1}}{\nu_{2}}t_{1}-\frac{\hat{k}_{1}\cdot r}{\nu_{2}}\right)\right\}\exp\left[-i\frac{\nu_{2}}{\nu_{1}}\omega\left(t-t_{1}\right)\right]\cdot\exp\left(i\omega t_{1}\right)\exp\left(-ik_{1}\cdot r\right)\\&+p_{2}A_{4}u\left(t-t_{1}-\frac{\hat{k}_{2}^{^{\prime}}\cdot r}{\nu_{2}}\right)\cdot\exp\left[i\left(\omega t-k_{2}\cdot r\right)\right]
	\end{aligned}
	\label{eq:fourty_six}
\end{equation}
\end{widetext}
where $q_{1}=z\cos\theta_{1}-x\sin\theta_{1}$, $p_{2}=z\cos\theta_{2}+x\sin\theta_{2}$.From Eq.~(\ref{eq:fourty_six}), we can observe that the first and second terms are the waves propagating in the forward and backward directions. And the third term is a reflected wave from the wave represented by the second term. The last term is a transmitted wave after $t=t_{1}$.

Here, we present Green’s functions in time varying media, including the exact closed form in homogeneous media with time interfaces and the general form in inhomogeneous media. We start by introducing a simple configuration [\cite{felsen1970wave}], which involves a nondispersive homogeneous medium whose wave velocity $v(t)$ changes rapidly from $v_{1}$ for $t<t_{1}$ to $v_{2}$ for $t<t_{2}$. For propagation at the direction z, the scalar Green’s function satisfies the wave equation
\begin{equation}
\left[\frac{\partial^{2}}{\partial z^{2}}-\frac{1}{\nu^{2}(t)}\frac{\partial^{2}}{\partial t^{2}}\right]G\left(z,z^{\prime};t,t^{\prime}\right)=-\delta\left(t-t^{\prime}\right)\delta\left(z-z^{\prime}\right)
\end{equation}
with the causality condition $G\equiv0$ for $\mathrm{t<t'}$ and the continuity conditions $G$, $\partial G/\partial t$ continuous at $t=t_{1}$. The solution for a spatially and temporally distributed source is obtained on multiplying $G$ by the source function $f(z^{\prime},t^{\prime})$ and integrating over the source domain $(z^{\prime},t)$. For $t<t_{1}$, the solution  $G_{0}$, is the same as in a time-invariant homogeneous medium with wave velocity $v_{1}$.
\begin{equation}
G_{0}=(\nu_{1}/2)U\left(\nu_{1}T-|Z|\right),T=t-t^{\prime},Z=z-z^{\prime}
\end{equation}
where the Heaviside unit function $U(a)$ equals l for $a>0$ and 0 for $a<0$. For $t>t_1$, the solution is represented as the form $F(\nu_{2}t\pm z)$
\begin{equation}
	\begin{aligned}
	G_{1}=(\nu_{1}/4)(1+\nu_{1}/\nu_{2})U(\nu_{2}\tau+\nu_{1}T_{0}-|Z|)\\ +(\nu_{1}/4)(1-\nu_{1}/\nu_{2})[U(\nu_{1}T_{0}-\nu_{2}\tau-|Z|)\\
	-U(\nu_{2}\tau-\nu_{1}T_{0}-|Z|)]
	\end{aligned}
\end{equation}
where $T_{0}=t_{0}-t^{\prime}$, $\tau=t-t_{0}^{\prime}$.

Note that the outgoing wavefront is cylindrical or spherical rather than plane, and interesting focusing phenomena arise because of collapse of the reflected wavefront toward its center. This is because we assume a point source. A response for a pulsed point source at $r=0$ satisfies the wave equation
\begin{equation}
	\left\{\nabla^{2}-\left[1/\nu^{2}(t)\right]\left(\partial^{2}/\partial t^{2}\right)\right\}G\left(r,t,t^{\prime}\right)=-\delta\left(t-t^{\prime}\right)\delta\left(r\right)
\end{equation}
for $t<t_{0}$, the Green’s function for the spherically expanding pulse can be formulated as
\begin{equation}
	G_{0}=\nu_{1}\delta\left(c_{0}T-r\right)/4\pi r,t<t_{0}
\end{equation}
for $t>t_{0}$, the Green’s function is expressed as
\begin{equation}
	\begin{aligned}&G_{1}=(\nu_{1}/8\pi r)\Big\{(1+\nu_{1}/\nu_{2}) \delta\big(\nu_{1}T_{0}+\nu_{2}\tau-r\big)\\&+(1-\nu_{1}/\nu_{2})\Big[\delta\big(\nu_{1}T_{0}-\nu_{2}\tau-r\big)-\delta\big(\nu_{2}\tau-\nu_{1}T_{0}-r\big)\Big]\}\end{aligned}
\end{equation}
It has been discussed how a forward and a backward wave will be produced after inducing a time interface. One will propagate in the original direction, and the other will propagate in the reverse direction with the original one [\cite{morgenthaler1958velocity}]. The frequency of these two waves has been shifted from $\omega_{0}$ to $\omega=\omega_{0}(1+\alpha^{2})^{\frac{1}{2}}$ [\cite{morgenthaler1958velocity}].

\subsection{Wave time space duality}
The space-time duality in acoustic equations has aroused interest in the realm of boundaries \cite{long2023time}. In this section, we explore the connection between scattering problems in the time-varying acoustic wave equation and the space-time duality.

\begin{figure}
	\centering	
	\includegraphics[height=6cm,width=8cm]{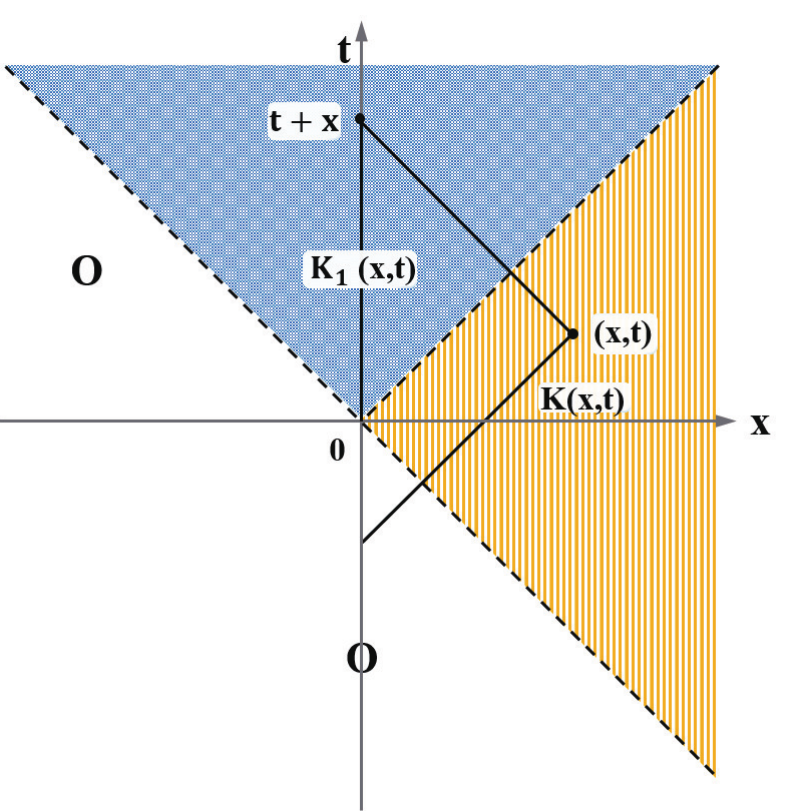}	
	\caption{\label{fig:2}The fundamental solutions of Eq.\ref{eq:fifty_three}.}	
\end{figure}

A common formulation of the acoustic wave equation can be defined as
\begin{equation}
	\left(\frac{\partial^{2}}{\partial z^{2}}-\frac{\partial^{2}}{v^{2}(z)\partial t^{2}}\right)u(z,t)=-s(t),
	\label{eq:fifty_three}
\end{equation}
where $v$ is the wave velocity, $u$ is the wavefield, $z$ is the distance. Similar to the standard wave propagation, the role of space and time are reversed. The time interface will lead to the wave splitting into FW and BW waves. From Eq.~(\ref{eq:fifty_three}), one can observe that it is totally symmetric to the spatial boundary condition. We now discuss a coupled first-order differential equation system for dual solution
\begin{equation}
	\frac{d}{dz}\left[\begin{array}{c}{u}\\{u_{z}}\end{array}\right]=\left[\begin{array}{cc}{0}&{1}\\{-\frac{\omega^{2}}{v^{2}(z)}}&{0}\end{array}\right]\left[\begin{array}{c}{u}\\{u_{z}}\end{array}\right].
\end{equation}
The dual solutions are defined as
\begin{equation}
	f=\delta(t-\zeta)+K(\zeta,t),
	\label{eq:fifity_five}
\end{equation}
where $K$ is the kernel of scattering. The equation governing dual fundamental solutions are written as
\begin{equation}
	0=K(\zeta,t)+R(t+\tau)+\int_{-\zeta}^{\zeta}K(\zeta,\tau)R(\zeta+\tau)d\tau,
\end{equation}
where $R$ is the reflection signal. Fig.\ref{fig:2} shows schematic of the fundamental solutions. The dual solution corresponds precisely to the solution of the inverse scattering integral equation. The dual solution of the above numerical simulation can be formally expressed as Eq.~(\ref{eq:fifity_five}). The reflection events after the first-arrival event-line (Diagonal line of x and t axis) can be observed, therefore, the area of the yellow triangle on the right represents the causal solution. In contrast, or in a complementary way, the blue area represents the “past area” recording the events in the past. In the future area, the events are observed by forward propagation, whereas in the past area, the events are encountered through back propagation (reverse-time). The causality means the role of time and space are interchanged.

\section{RESULTS}
To understand the characteristics of acoustic wave propagation with time interfaces, we solved Eq.~(\ref{eq:thirty-four}) using the staggered-grid Finite Difference Time Domain (FDTD) scheme. First, we compare and analyze the wave field characteristics under different time interfaces in homogeneous models. Then, we consider a spatial interface between two media where one region is modulated using a time interface to verify the effects of the time interface. Finally, we incorporate the temporal interface equations into the BP model [\cite{huang2023full}] to investigate the properties of wave propagation.
\subsection{Homogeneous media with time interfaces}
\begin{figure}
	\includegraphics[width=3.6in]{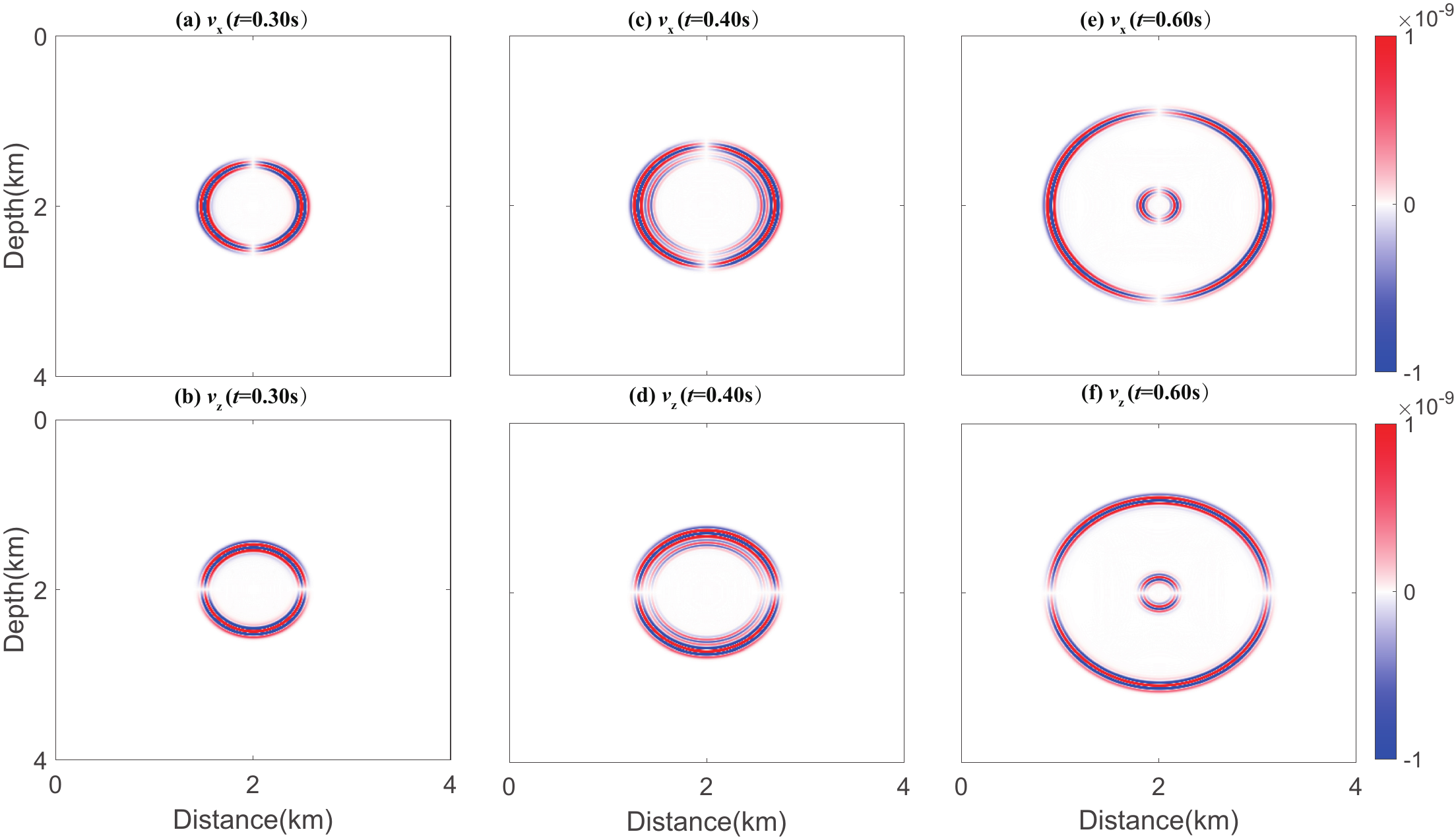}
	\caption{\label{fig:3}Snapshots of the particle velocity for $v_{x}$ and $v_{z}$. (a-b) Schematic representation of the velocity of the medium before it is modified in time. A wave generated by an initial Ricker wavelet is traveling in a medium with $v$ = 2 km/s, and a temporal interface is applied at $t$ = $t_{1}$ = 0.37 s. (c-f) particle wave velocity at a time $t$ = 0.4 s and $t$ = 0.6 s after the time interface is applied.}
\end{figure}

Let us first consider the case of acoustic wave scattering with different time interfaces in an homogeneous, unbounded medium where the wave propagates. For times  the initial velocity within the whole medium is set to $v = 2 $ km/s. Then, at $t=t_{1}=0.37$ s, the velocity of the whole medium is changed to a value of $v_{1}$ = 8944 m/s, the coefficient is $\alpha$ = 0.95 in Eq.~(\ref{eq:thirty-four}). The schematic representation of this configuration is shown in Figs. \ref{fig:3}a-b. The simulation space consists of a square region of size 4 km along both the x and y directions. The simulation step is divided into small squares of size 5 m and a time step of 0.0001 s is implemented. These parameters are then used in Eq.~(\ref{eq:thirty-four}) to calculate the scattered fields for times before and after the time interface is applied. The Ricker wavelet [\cite{ricker1943further,ricker1944wavelet}] is implemented with a central frequency of 25 Hz. And the source is located at the position (x = z = 2 km) and the receiver is positioned at (x=1.8 km, z=1.9 km).

\begin{figure}
	\includegraphics[width=3.6in]{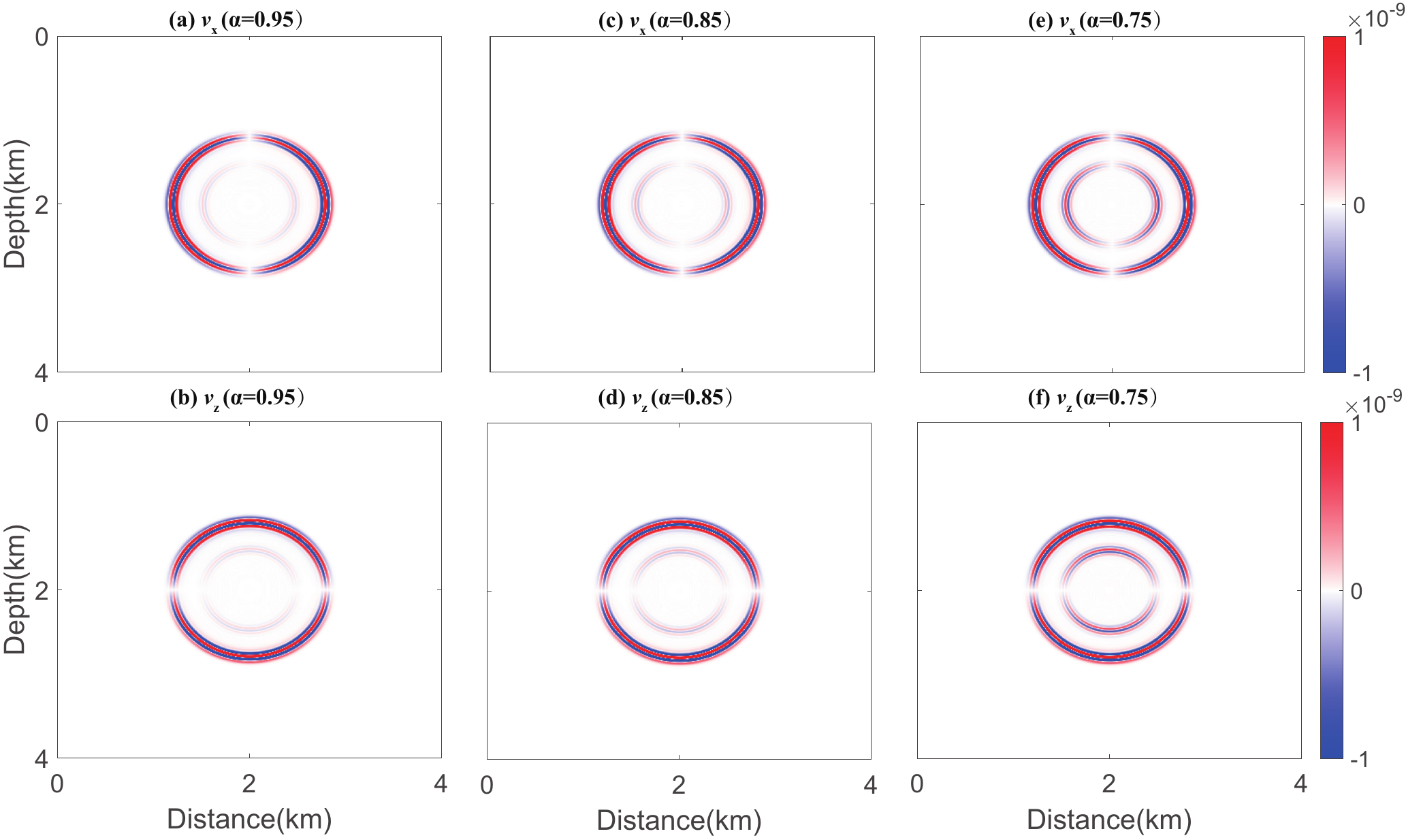}
	\caption{\label{fig:4}Snapshots of the particle velocity for $v_{x}$ and $v_{z}$ of $\alpha$ = 0.95, 0.85, 0.75. Results of acoustic wave pulses scattered at time interfaces with a single carrier Ricker wavelet input pulse. The BW and FW are created within induced temporal interface depending on the values where applying different temporal interfaces of different values. The value for (a), (b) is 0.95, for (c), (d) is 0.85 and for (e), (f) is 0.75.}
\end{figure}

\begin{figure}
	\centering	
	\includegraphics[width=3.5in]{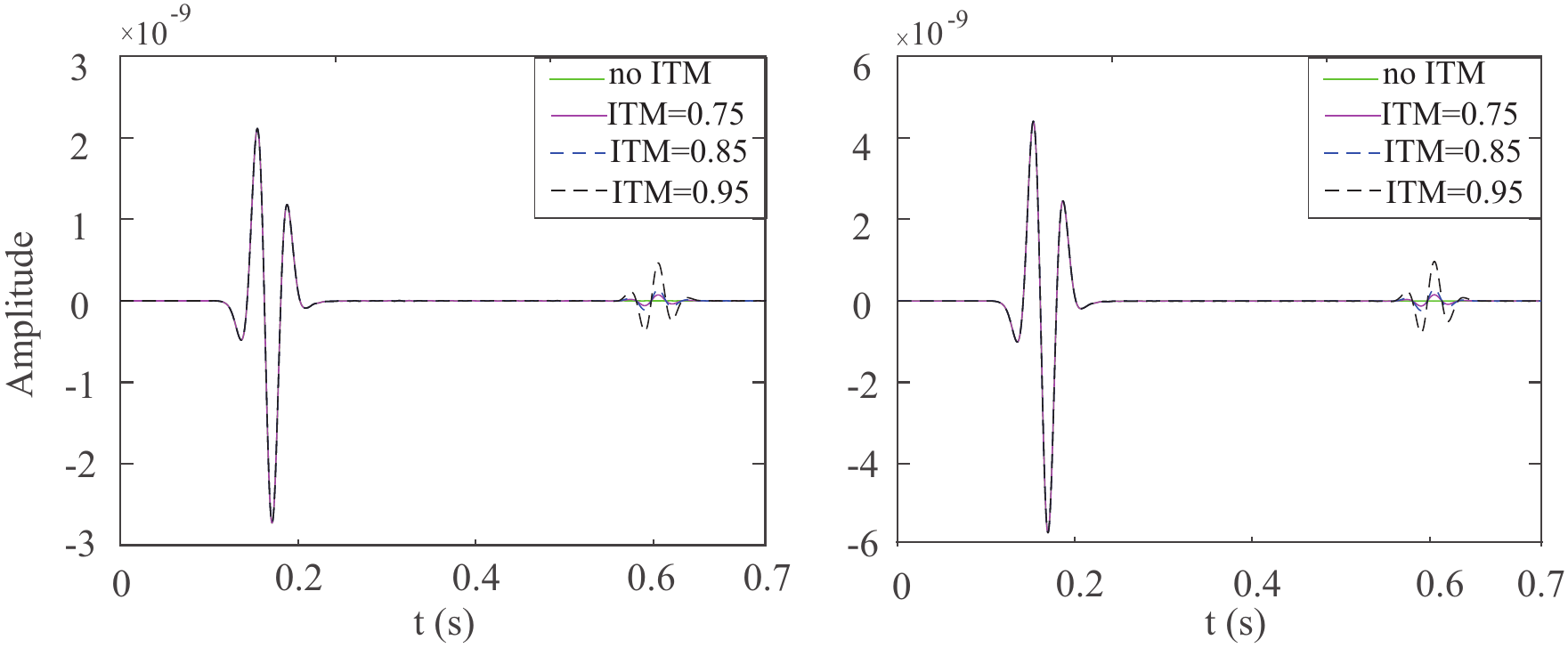}	
	\caption{\label{fig:5}The synthetic recordings are made at the receiving point with TI perturbation coefficients of $\alpha$ = 0.95, 0.85, 0.75. At $t$ = 0.37 s, a temporal interface is applied to the medium. The amplitudes of the FW and BW waves are larger in time interfaces where $\alpha$ = 0.95, 0.85, 0.75 are changed to larger values. (a) is the wave amplitude of x direction, (b) is the wave amplitude of y direction.}	
\end{figure}

With this configuration, snapshots of the particle velocity wavefields for $v_{x}$ and $v_{z}$ at different times are shown in Fig.\ref{fig:3}, where an acoustic wave is traveling within a medium. At $t=t_{1}$=0.37 s, a temporal interface is applied to the medium. As explained above, larger/smaller values of $\alpha$ mean that the perturbation to the velocity of the medium is smaller/larger, respectively. As shown in Figs. \ref{fig:3}c-d, two waves are created after inducing the time interface: one traveling forwards (diverging from the source) and one backwards (converging towards the source), noting that the generated FW and BW waves have different amplitudes. To further investigate the impact of various velocity disturbance intensities on wave propagation characteristics, we compared the snapshots of the particle velocity wavefield (see Fig. \ref{fig:4}) and synthetic recordings at the receiving point (see Fig. \ref{fig:5}) with time interface perturbation coefficients of $\alpha$ = 0.95, 0.85, 0.75 in Eq.~(\ref{eq:thirty-four}). As observed in these figures, the perturbation generated by the time interface has some effect on the energy and propagation of the primary wave (P wave), and the amplitude of the time inference (TI) wave significantly decreases as the TI coefficient decreases. The BW wave simply travels toward the source direction and refocuses the source. The FW wave continue its propagation in the direction of the initial wave, and when it reaches the spatial interface, spatial transmission and reflection of the wave occurs.

\begin{figure}
	\includegraphics[width=3.6in]{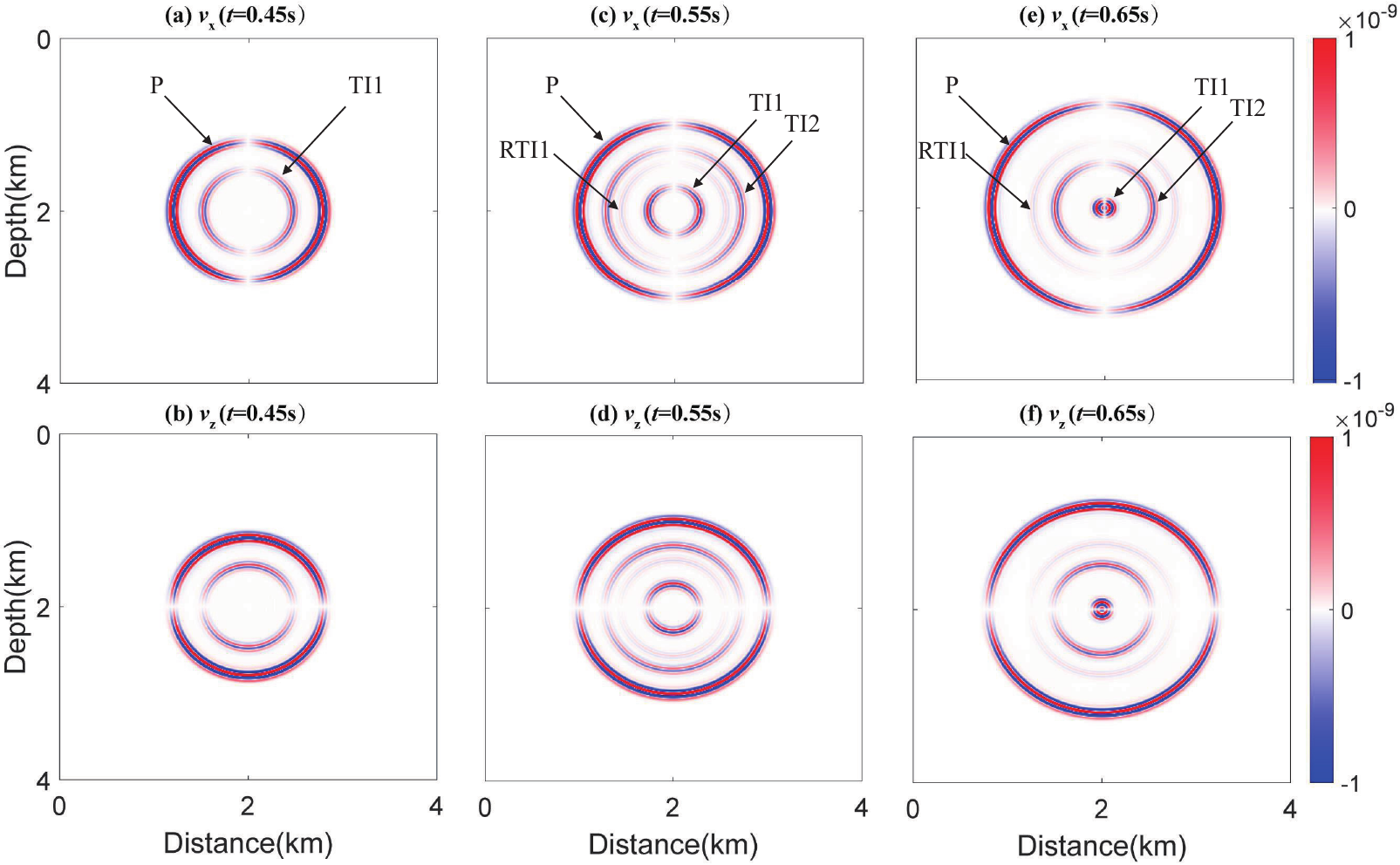}
	\caption{\label{fig:6}Snapshots of the particle velocity for $v_{x}$ and $v_{z}$ with two time interfaces set at $t$ = 0.37 s and $t$ = 0.48 s. The TI waves generated by the temporal interface at $t$ = 0.42 s. (a) and (b) are the particle velocity for $v_{x}$ and $v_{z}$ at $t$ = 0.45 s, (c) and (d) are the particle velocity for $v_{x}$ and $v_{z}$ at $t$ = 0.55 s, (e) and (f) are the particle velocity for $v_{x}$ and $v_{z}$ at $t$ = 0.65 s. }
\end{figure}

We have also performed numerical tests when considering a system with two time interfaces (as shown in Fig. \ref{fig:6}), two spatial interfaces are induced at the same time using time interfaces with different values. Fig. \ref{fig:6} shows snapshots of the particle velocity wavefield for $v_{x}$ and  $v_{z}$, with two time interfaces set at $t$ = 0.37 s and $t$ = 0.48 s, respectively. In this case, the velocity is changed from 2000 m/s to 8944 m/s. Compared to the two types of waves at $t$ = 0.45 s, the snapshots at $t$ = 0.55 s and 0.65s contain forward and backward waves: the P wave, the TI 1 wave perturbed by the first-time interface at $t$ = 0.37 s, the TI 2 wave perturbed by the second-time interface at $t$ = 0.48 s, and the perturbation wave RTI1 of TI 1. These successive time reflection events lead to temporal wave interference. The scattering behavior within a temporal slab differ from that of a spatial slab. While multiple scatterings within a spatial slab lead to a superposition of refracted and reflected waves, a temporal slab generates four distinct scattered waves following the second time interface.
\subsection{Layered model with time interface}
\begin{figure}
	\includegraphics[width=3.5in]{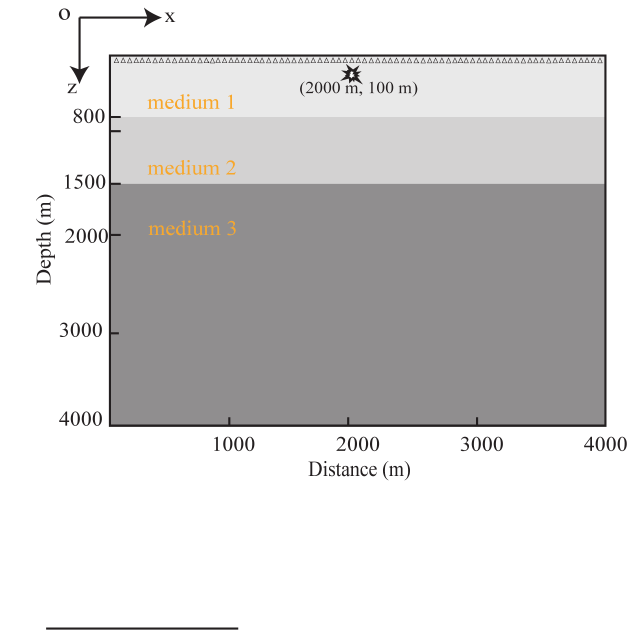}
	\caption{\label{fig:7}The layered model. The values of velocity for three layers are 2000 m/s, 3000 m/s and 4500 m/s, respectively. The velocity in the entire medium is changed to 8944 m/s, 13416 m/s and 20124 m/s.}	
\end{figure}

To analyze the propagation characteristics of the temporal interface wave response, we have constructed a three-layer model with the spatial interface located at a depth of z = 0.8 km and 1.5 km, which is shown in Fig. \ref{fig:7}. The velocity is set at $v_{0}$ =2000 m/s above the interface of 0.8 km, 3000 m/s below the interface of 1.5 km, and 4500 m/s for others. We consider the velocity 2000 m/s, 3000 m/s and 4500 m/s for the layered model within the entire space for times (shown in Fig. \ref{fig:7}), then at time $t$ = 0.48 s, the velocity in the entire medium is changed to 8944 m/s, 13416 m/s and 20124 m/s. The source is applied at the designated position (2 km, 0.1 km), the time interface set at $t$ = 0.48 s with time interface perturbation coefficients of $\alpha$ = 0.95, and other parameter settings are the same as those in the homogeneous model.

\begin{figure}
 	\includegraphics[width=3.7in]{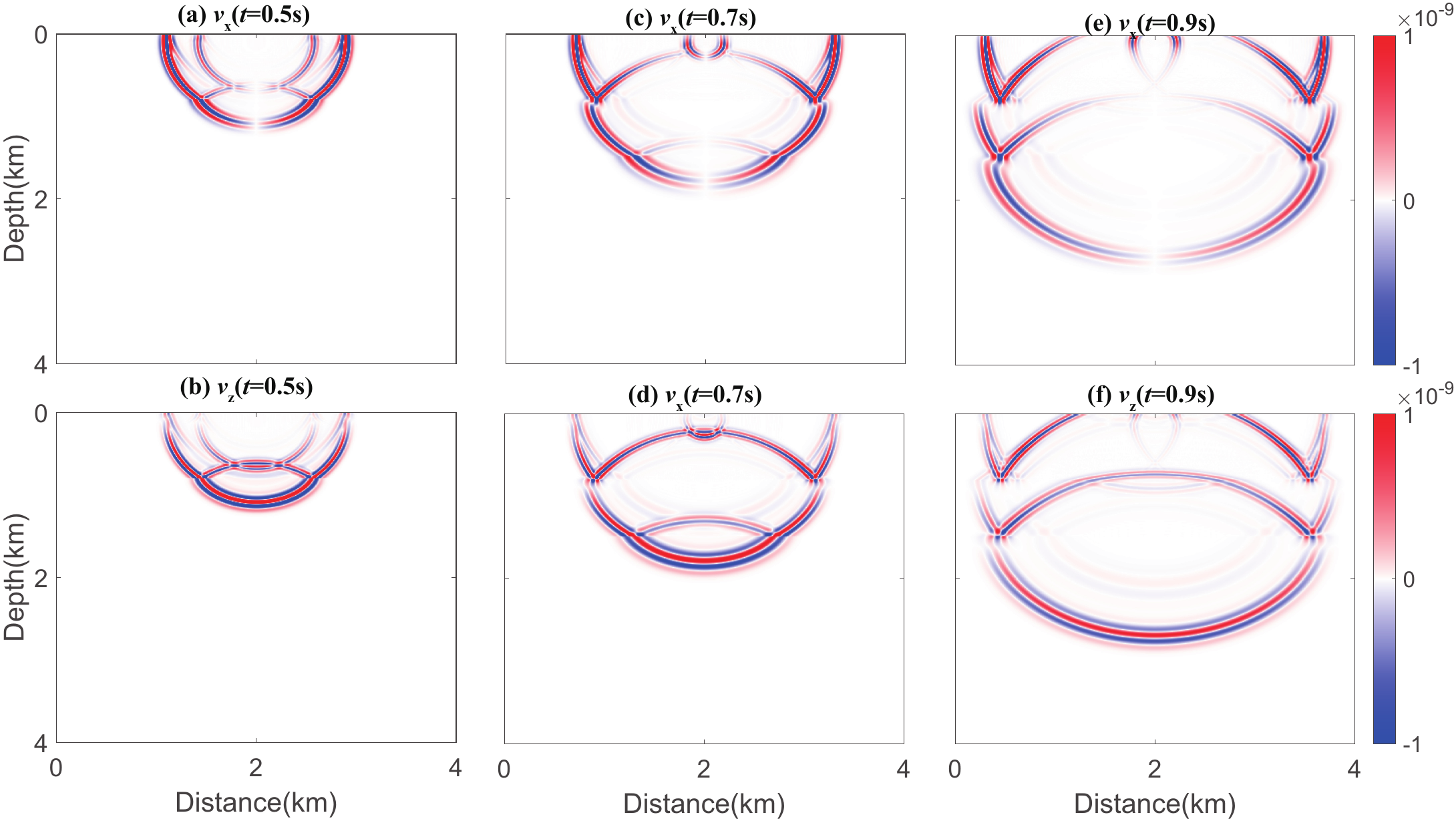}
 	\caption{\label{fig:8}Snapshots of the particle velocity for $v_{x}$ and $v_{z}$  at 0.5s, 0.7s and 0.9s. }
\end{figure}

\begin{figure}
 	\includegraphics[width=3.8in]{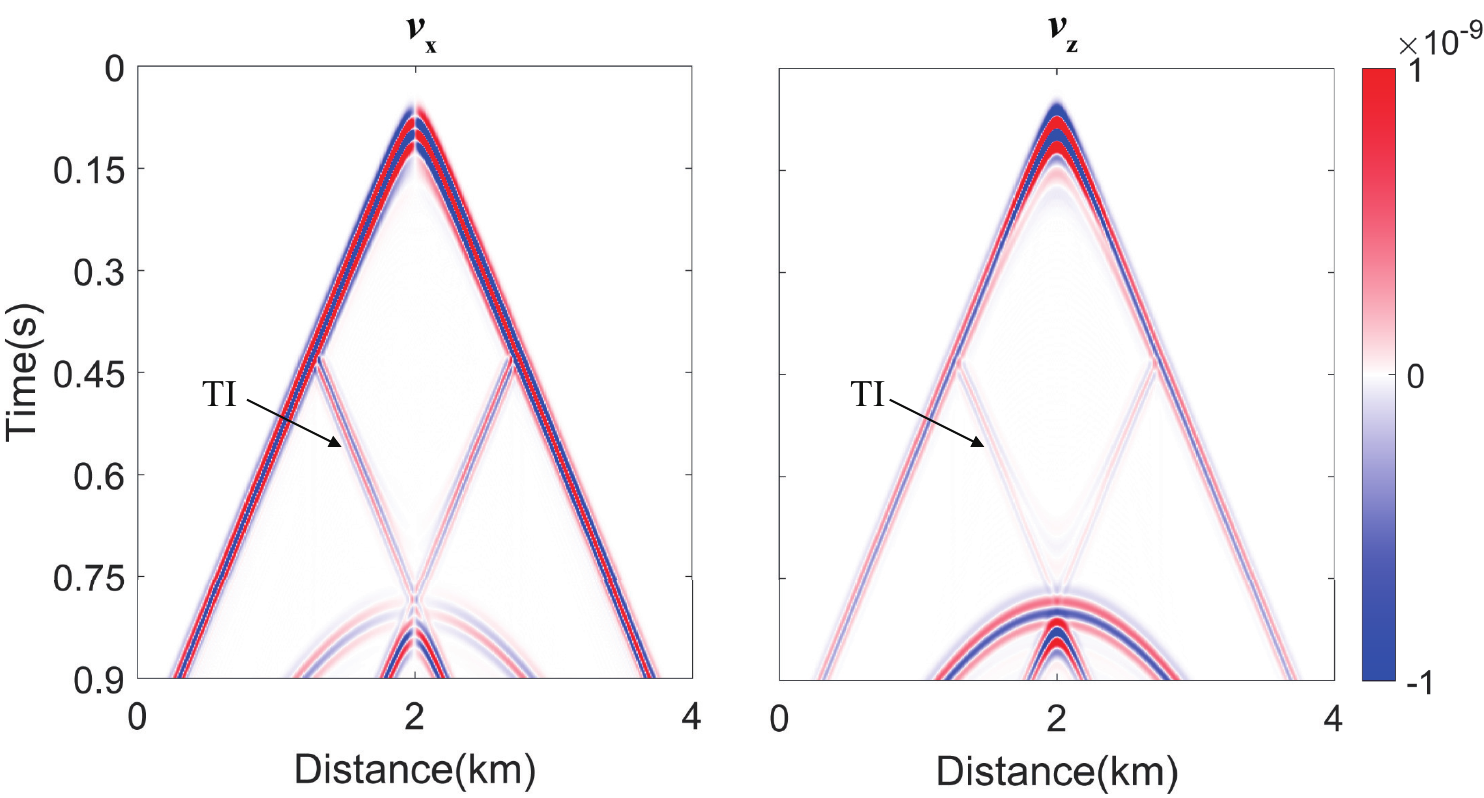}
 	\caption{\label{fig:9}Time domain seismic record of the particle velocity for $v_{x}$ and $v_{z}$. Amplitudes of the total time scattering signals are a function of receivers. Output response referenced at the input Ricker wavelet when adjusting the triggering of the time interface. Demonstration of a spatial reflection and a time reflection with the time interface happening at $t$ = 0.42 s.}
\end{figure}

Fig. \ref{fig:8} shows the snapshots of the particle velocity wavefields for $v_{x}$ and $v_{z}$ at 0.5s, 0.7s and 0.9s. The TI waves generated by the temporal interface at $t$ = 0.42 s, as well as the direct, reflected, and transmitted waves at the layer interfaces at z = 0.8 km and 1.5 km, can all be clearly observed. It illustrates that after 0.42 seconds, two waves are created, one propagating in a forward (diverging from the source) and backward direction (converging towards the source), due to the BW and FW at the temporal discontinuity. On the other hand, when the FW wave reaches the spatial boundary, one wave is transmitted into layer three, one is reflected with forward direction (diverging from the source). Seismic records of the particle velocity for $v_{x}$ and $v_{z}$ are displayed in Fig. \ref{fig:9}. The receivers were placed at a depth of z = 100 m and recorded all the direct and reflected seismic waves, including the TI wave.The direct and reflected waves due to spatial interface are stronger than the FW and BW wavefield produced along the induced time interface. The time reflected wave and time-refracted wave in time domain show amplitude change induced by the time interface. The amplitudes of these waves depend on the extent of the velocity change. The FW wavefield produced along the induced interface is stronger.
\subsection{BP model with time interfaces}
\begin{figure}
	\includegraphics[width=3.5in]{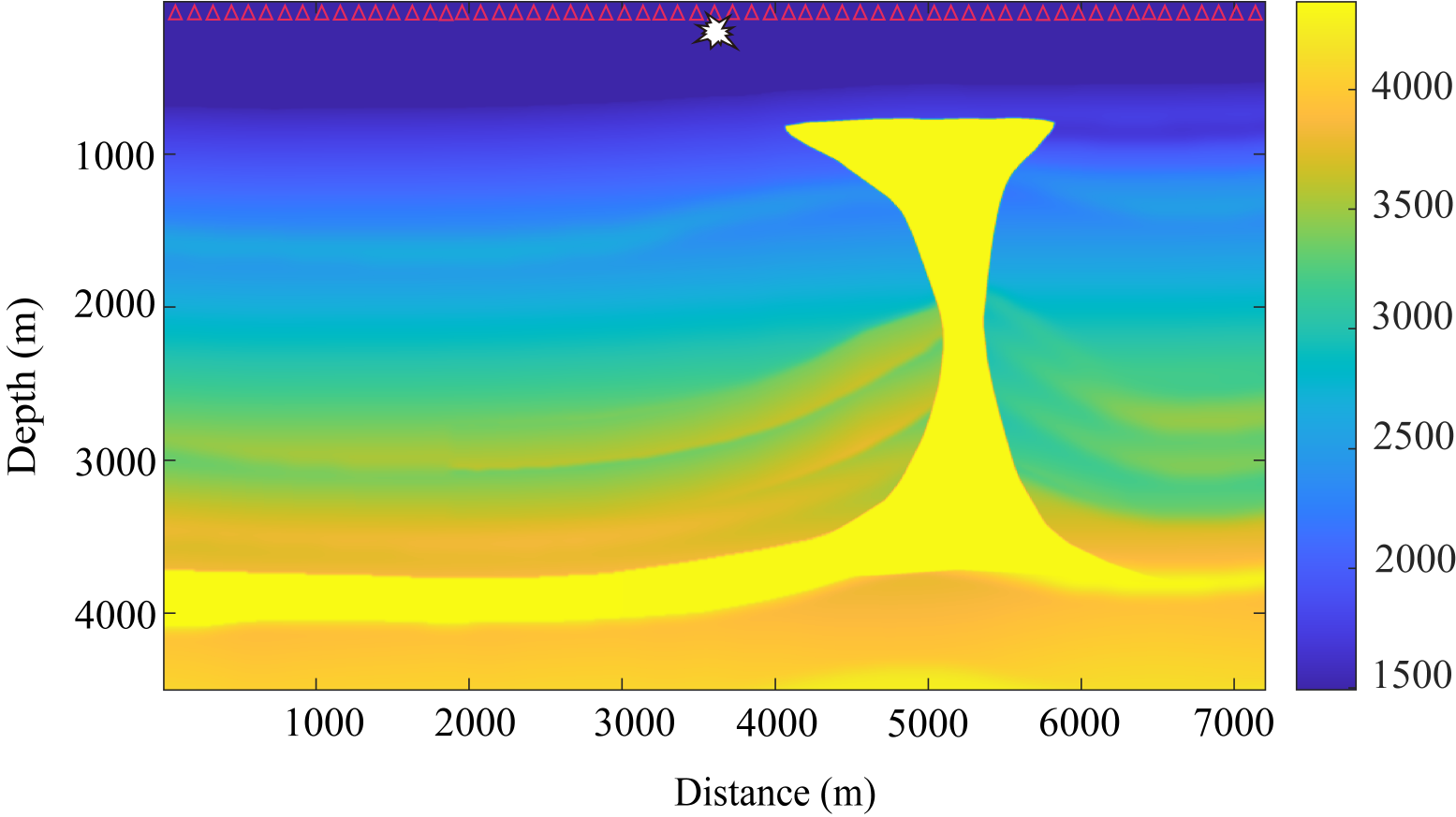}
	\caption{\label{fig:10}The modified BP model.There is a high velocity within the model where the wavefields are complicated. The velocities of the waves are heterogeneous within the model.}
\end{figure}

\begin{figure}
	\includegraphics[width=3.6in]{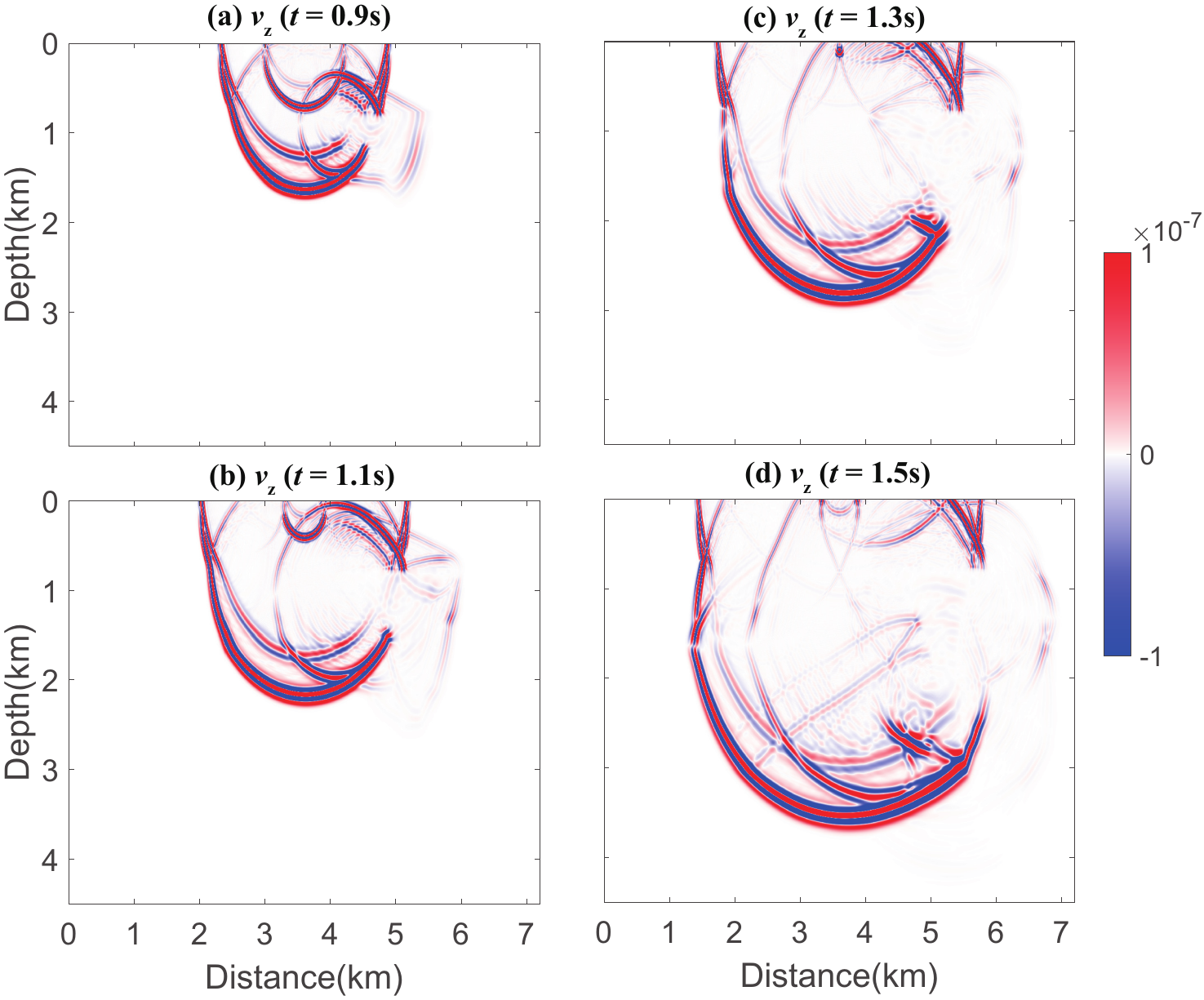}
	\caption{\label{fig:11}Snapshots of the particle velocity for $v_{z}$ at 0.9, 1.1, 1.3, and 1.5 s. The TI waves generated by the temporal interface at $t$ = 0.68 s. (a) is the particle velocity of wavefield for $v_{z}$ at $t$ = 0.9 s, (b) is the particle velocity of wavefield for $v_{z}$ at $t$ = 1.1 s, (c) is the particle velocity for   at  =1.3 s, (d) is the particle velocity of wavefield for $v_{z}$ at $t$ = 1.5 s.}
\end{figure}

\begin{figure}
	\includegraphics[scale=0.4]{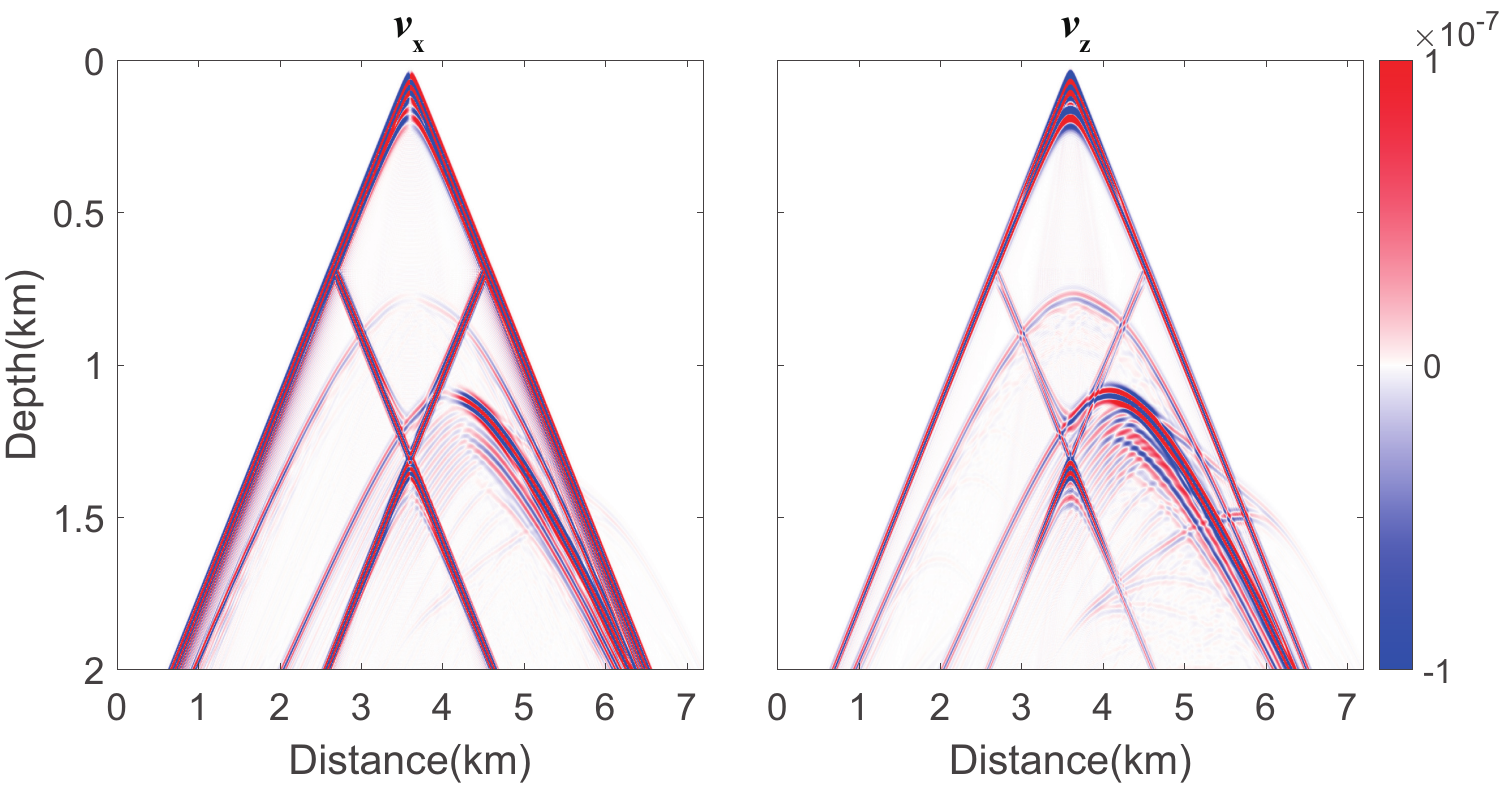}
	\caption{\label{fig:12}The particle velocity seismic records of the modified BP model for $v_{x}$ and $v_{z}$. Corresponding measurements with multiple receivers and simulations of the waves at the surface as a function of time. The undesired portion of the incoming signal is reflected with time interface refer to signal measurements at the surface. The receivers were placed at a depth of z = 100 m and are evenly distributed across the surface, having a range of 0-7.2 km. The receivers recorded all the direct and reflected waves, including the TI waves. The FW and BW are created by time interface. The reflected waves are created by spatial interfaces in which the velocities of the waves are different. The time reflected wave and time-refracted wave in time domain, showing amplitude change induced by the time interface. Demonstration of a spatial reflection and a time reflection with the time interface occurring at $t$ = 0.68 s.}
\end{figure}

To further validate the numerical implementation and effectiveness of the temporal interface acoustic wave equation, we used a modified BP model \cite{billette20052004} to examine the wave propagation characteristics, as shown in Fig. \ref{fig:10}. The methods are numerically investigated for different configurations, including the integration of temporal boundary and spatial interfaces, the combination of temporal interfaces with spatial interfaces, and the impact of both temporal and spatial interfaces. To facilitate the investigation of wave scattering with time interfaces, especially in the presence of multiple such temporal interfaces, we use such complicated velocity model. The source is applied at the position (3.6 km, 2.25 km) with the time interface set at $t$ = 0.68 s. The model size is 7.2 × 4.5km, with 100 grid points for the perfectly matched layer (PML). The velocity in the entire medium is changed to 4.47 times of the original velocity. Figure 11 represents snapshots of the particle velocity for $v_{x}$ at 0.9, 1.1, 1.3, and 1.5 seconds. They clearly show both the incident wave and the reflected and transmitted waves. We also compute seismograms of particle velocities of waves for the modified BP model (see Fig. \ref{fig:12}), where all the temporal and spatial waves produced by the temporal and induced spatial interfaces can be observed. The receivers are evenly distributed across the surface, having a range of 0-7.2 km. Waves in such a model provides wave propagation with the combination of the spatial and temporal inhomogeneities. It is demonstrated numerically how such a temporal boundary can produce FW and BW in the medium with complicated spatial inhomogeneities. From Fig. \ref{fig:11}, one can observe that two waves are created, one propagating in a forward and backward direction, due to the time interface. The waves will travel towards the until all converge. On the other hand, each spatial interface within the model will create transmitted and reflected wavefields which will travel towards the diverging directions. The waves due to the spatial interfaces will split into two waves when temporal interface occur as well. This means that the interaction of the FW with the spatial boundary will create two more waves: one transmitted, with a new wavelength as it will travel into other locations, and one reflected (due to the spatial boundary). Note that the direct and reflected waves due to spatial interface are stronger than the FW and BW wavefield produced along the induced time interface. This study provides information on how the waves propagates when spatial and temporal interfaces are combined in a model and what physical phenomena are if spatial interfaces are induced by applying temporal interfaces to spatial inhomogeneous medium.
\section{CONCLUSIONS}
Wave time scattering problems are difficult to solve for time interfaces with  time-varying media. With temporal interfaces over very short timescales, it is hard to obtain solutions of the time perturbation wave equation. In this work,  we have derived formulations of the Lippmann-Schwinger integral equations for acoustic wave scattering with time interfaces, we introduce wave reflection, Green's functions, frequency translation and time space duality. In particular, we present numerical solutions of the time interface acoustic wave equation, even in the absence of strongly scattering media. We have demonstrated the capability of the finite difference method in effectively solving the time perturbation wave equation for  scattering problems that can be applied for wave focusing, negative refraction, wave control and manipulation, broadband spectral filtering and frequency. We have shown that the framework is able to accurately estimate the wavefields with a time interface.

We have employed the perturbation theory to derive the Lippmann-Schwinger integral equations. We apply the finite difference  approach to solve time perturbation wave equation that recast strongly nonlinear scattering problems into a globally linear framework with wave physics. Our approach is designed to obtain wavefields after inserting a time interface, which provides wave refocusing, and have been difficult to realize using alternative methods. Building on the time perturbation wave equation, we present the Green's functions for focusing wavefields where all the scattered wavefields have approximately the same duration as the input wavefields. We have shown that a temporal time slab produces a total of four scattered waves after the second time interface. Throughout this paper, we have used physical insight from the time perturbation wave equation. We have demonstrate time interfaces  in a large scale Earth structure.

There are many ongoing challenges and promising directions that motivates future work. First, it has been recognized that by implementing such time interfaces, we form a temporal slab in which the reflected and refracted wavefields emanate at the time interface. The presented focusing waves establish the fundamental building blocks to exploit time as an additional degree of freedom for wave focusing. Further, there is ongoing work to improve extreme wave manipulation in geophysical applications. This work attempts to promote wave focusing by implementing instantaneous time mirrors. Continued effort will be required for efficient time reversal. There is a tremendous opportunity to open a pathway to employ time interfaces for broadband, efficient phase conjugation and frequency conversion arising over very short timescales. It will be important to extend these methods to  applications in electromagnetics, elastic waves, acoustics, seismics and photonics \citep{huang2023full}. Our method can be applied to inverse scattering problems. The role of scattering theory has dramatically increased over the past two decades due to it is theoretical foundation for seismic inversion but still suffers from the ill-posed nature of inversion and nonlinearity problems. Currently, many full waveform inversion methods operate scattering integral approaches and in the near future inverse scattering will be an important part of seismic inversion.

\appendix

\section{Dual Solution}
This section gives a derivation of the dual solution. In the space-time domain, the pressure wavefields satisfies the scalar wave equation
\begin{equation}
	\left (\bigtriangledown ^{2} -\frac{1}{c^{2}} \frac{\partial^2 }{\partial t^2}\right )u\left ( {\bf x} ,t \right )=s\left ( t \right ).
\end{equation}
We assume $V=\frac{1}{c^{2}}-\frac{1}{c^{2}_{0}}$, we obtain
\begin{equation}
	\left ( \bigtriangledown ^{2} -\frac{1}{c_{0}^{2}}\frac{\partial^2 }{\partial t^2}-V\left ( \mathbf{x} \right )\frac{\partial^2 }{\partial t^2}\right )u\left ( \mathbf{x,t} \right )=s\left ( t \right ) .
\end{equation}
When we set the operators
\begin{equation}
	L=\left ( \bigtriangledown ^{2}-\frac{1}{c^{2}} \frac{\partial^2 }{\partial t^2}\right ),
\end{equation}
and
\begin{equation}
	L_{0}=\left ( \bigtriangledown ^{2}-\frac{1}{c^{2}_{0}} \frac{\partial^2 }{\partial t^2}\right ) ,
\end{equation}
then the scattering potential is as follows
\begin{equation}
	L-L_{0}=\left ( \frac{1}{c^{2}}-\frac{1}{c^2_{0}}\right ) \frac{\partial^2 }{\partial t^2}=V\frac{\partial^2 }{\partial t^2} .
\end{equation}
The scattering solution of Eq.~ (\ref{eq:fifty_three}) is
\begin{equation}
\begin{aligned}
u\left(t,{\bf x} \right)=u_{0}\left ( t,{\bf x} \right )+\int d^{3}dt^{{}'}g_{0}\left ( t-t^{{}'} ,{\bf x}\right )V \frac{\partial^2 }{\partial t{}'^2}u\left ( t,{\bf x}^{{}'} \right ),
\label{eq；A6}
\end{aligned}
\end{equation}
where the Green's functions are defined by
\begin{equation}
	\left ( \Delta -\frac{\partial^2 }{\partial t^2} \right )g_{0}^{\pm }\left ( t-t{}' ,{\bf x},y\right )=\delta ^{3}\left ( {\bf x}-y \right )\delta \left ( t-t{}' \right ),
\end{equation}
and
\begin{equation}
	\left ( \Delta -\frac{\partial^2 }{\partial t^2}-v\left ( {\bf x} \right ) \right )g^{\pm }\left ( t-t{}' ,{\bf x},y\right )=\delta ^{3}\left ( {\bf x}-y \right )\delta \left ( t-t{}' \right ).
\end{equation}
The Green's functions have the relation as
\begin{equation}
	\begin{aligned}
		g^{\pm }\left ( t-t{}',{\bf x},y \right ) &= g_{0}^{\pm }\left ( t-t{}' ,{\bf x},y\right ) \\
		&+ \int d^{3}{\bf x}{}''dt{}''g^{\pm }\left ( t-t{}'',{\bf x},{\bf x}{}''\right )\\
		&\times  V \left ( {\bf x}^{{}''} \right ) \frac{\partial^2 }{\partial t^2}g_{0}^{\pm }\left ( t{}''-t{}',{\bf x}{}'' ,{\bf x} \right ),
	\end{aligned}
\end{equation}
\begin{equation}
	u^{-}\left ( t,1,{\bf x} \right )=u^{+}\left ( -t,-1,{\bf x} \right ).
	\label{eq:A10}
\end{equation}

Eq.~(\ref{eq:A10}) is the basis for wave equation type inverse scattering problems. Causality leads to the physical interpretation of the solutions of the equation.  The wave fields $u^{+}$ is zero in the region of $t<x$ whereas the wave fields $u^{-}$ is zero in the region of $t>x$. In addition, the Green's functions $G_{0}^{+ }\left ( t-t^{{}'}\right)$ is zero if $t-t{}'<x-x{}'$. Also, the Green's functions $G_{0}$ is zero if $t-t{}'>x-x{}'$. 

The quantum scattering theory is governed by the time independent \emph{Schr\"odinger equation} 
\begin{equation}
	\left ( \bigtriangledown ^{2}+k^{2} \right )\psi \left ( {\bf x},k \right )=V\left ( {\bf x} \right )\psi \left ( {\bf x},k \right ),
\end{equation}
where $\bigtriangledown ^{2} $ is the the Laplacian operator and $V$ is the scattering potential, which is a real-valued function, $u$ denotes the wave field, ${\bf x}$ denotes the space coordinate, while $k$ denotes the wave number.
If the incident signal is a plane wave, then the solution of the equation is determined by the Lippmann-Schwinger equation \cite{rose2001single}
\begin{equation}
	u^{+}\left ( k,1,{\bf x} \right )=e^{ik{\bf x}}+ \int_{-\infty }^{\infty }d y G_{0}^{+}\left ( k,\left | {\bf x}-y \right | \right )V u^{+}\left ( k,1,y \right ),
	\label{eq:A12}
\end{equation}
with a noninteracting Green's function
\begin{equation}
	G_{0}^{+}\left ( k, r \right )=- \frac{1}{4\pi} e^{\pm ikr}.
	\label{eq:A13}
\end{equation}
where $r=\left | {\bf x}-y \right |$, sign $+$ corresponds to the outgoing radiation condition, while the sign $-$ corresponds to the incoming radiation condition, and
\begin{equation}
	u^{+}\left ( k,1,{\bf x} \right )=u^{+*}\left ( -k,1,{\bf x} \right ),
	\label{eq:A14}
\end{equation}
where $*$ indicates complex conjugate. 

The time independent \emph{Schr\"odinger equation} can be related to equations in the time domain. We use the Fourier transform
\begin{equation}
	f \left ( t \right )=\left ( 2\pi  \right )^{-1}\int_{-\infty }^{\infty }\mathrm{exp}\left ( -ikt \right )f\left ( k \right ) dk,
\end{equation}
and transform the Schr\"odinger equation with respect to the wave number into the \emph{plasma wave equation}
\begin{equation}
	\left ( \frac{\partial^2 }{\partial {\bf x}^2}-\frac{\partial^2 }{\partial t^2} \right )u\left ( {\bf x},t \right )=V\left ( {\bf x} \right )u\left ( {\bf x},t \right ).
	\label{eq:A16}
\end{equation}
When the potential $V$ is zero, then Eq. (\ref{eq:A12}) becomes the acoustic wave equation in free space. One can observe that the plasma wave equation is a dispersive hyperbolic wave equation that propagates at velocity of 1. Note that there is no any physical interpretation for the variable $t$.

The solutions of the plasma wave equation is governed by the formal analog of the \emph{Lippmann-Schwinger equation}
\begin{equation}
	\begin{aligned}
	u^{\pm }\left ( t,1,{\bf x} \right )=\delta \left ( t-{\bf x} \right )\\
	+\iint G_{0}^{\pm }\left ( t-t^{{}'} ,r\right ) Vu^{\pm }\left ( t^{{}'} ,1,y\right )dt^{{}'}dy,
	\end{aligned}
	\label{eq:A17}	
\end{equation}
where
$G_{0}^{\pm }\left ( t, r \right )$ is the Green's function in the background medium,
and
\begin{equation}
	u^{\pm }\left ( t,1,{\bf x} \right )=\left ( 2\pi  \right )^{-1} \int_{-\infty }^{\infty }\textrm{exp}\left ( -ikt \right ) u^{\pm }\left ( k,1,{\bf x} \right )dk.
\end{equation}
The outgoing solution $u^{+}$ and the incoming solution $u^{-}$ are related by
\begin{equation}
	u^{-}\left ( t,1,{\bf x} \right )=u^{+}\left ( -t,-1,{\bf x} \right ).
	\label{eq:A19}
\end{equation}

Eq.~(\ref{eq:A13}) is the basis for Schrodinger type inverse scattering problems. Causality leads to the physical interpretation of the solutions of the equation.  The wavefield $u^{+}$ is zero in the region of $t<x$ whereas the wavefield $u^{-}$ is zero in the region of $t>x$. Also, the Green's functions $G_{0}^{+ }\left ( t-t^{{}'}\right)$ is zero if $t-t{}'<x-x{}'$. Then the Green's functions propagate from the past to the future. Further, the  Green's functions propagate from the future to the past meaning that the Green's functions $G_{0}$ is zero if $t-t{}'>x-x{}'$. 

The scattering potential for acoustic wave scattering problems is $V$ but the scattering potential for plasma wave scattering problems is $V\frac{\partial^2 }{\partial t^2}$. It should be noted that the Fourier transform of $\frac{\partial^2 }{\partial t^2}$ is $k^{2}$, which is frequency dependent. That makes the scattering potential stronger at high frequency. 

In this section, we outline the methodology for the fundamental solutions of the plasma wave equation, including the rightward travelling waves, leftward waves. We focus on the implementation and physical interpretation of the solutions and we discuss the algorithm used in the application. If readers don't want to go into the details of the mathematical derivations, they can understand that if we know the reflectivity of the medium, we can get the wavefields in the direction of time. The solution $u\left ( {\bf x},t \right )$ of Eq. (\ref{eq；A6}) can be written as the superposition of rightward travelling and leftward travelling waves
\begin{equation}
	u\left ( {\bf x},t \right )=f\left ( {\bf x}-t \right )+g\left ( {\bf x}+t \right ).
\end{equation}

In the region ${\bf x} \leq 0$, $V\left ( {\bf x} \right )=0$, then
\begin{equation}
	u\left ( {\bf x},t \right )=f_{-} \left ( {\bf x}-t \right )+g_{-}\left ( {\bf x}+t \right ).
	\label{eq:A21}
\end{equation}
If  we consider the solution $u\left ( {\bf x},t \right )$ with only the backward wave in ${\bf x}\leq 0$, then 
\begin{equation}
	u\left ( {\bf x} \right )=u\left ( t \right )=g_{-}\left ( {\bf x}+t \right ),
	\label{eq:A22}
\end{equation}
with
\begin{equation}
	f_{-}\left ( {\bf x}+t \right )=0, {\bf x} \leqslant 0.
	\label{eq:A23}
\end{equation}
Particularly, when ${\bf x}=0$
\begin{equation}
	u_{t}\left ( 0,t \right )-u_{{\bf x}}\left ( 0, t \right )=0.
\end{equation}

\begin{figure}
	\centering
	\includegraphics[height=6cm,width=8cm]{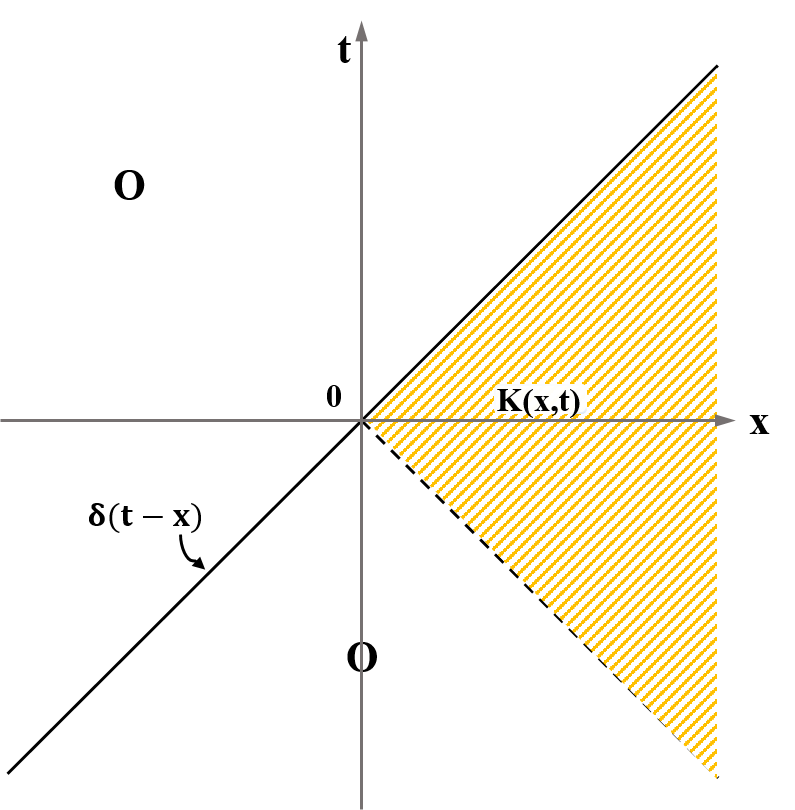}
	\caption{\label{fig.13}The anti-causal  solutions $f\left ( {\bf x} ,t \right )=\delta \left (t- {\bf x}  \right )+K\left ( {\bf x} ,t \right )$.}
	
\end{figure}

In the region ${\bf x} \geqslant 0$
\begin{equation}
	u\left ( {\bf x},t \right )=f_{+} \left ( {\bf x}-t \right )+g_{+}\left ( {\bf x}+t \right ).
\end{equation}
Following Tao \cite{tao2009israel},  we introduce 
\begin{equation}
	f\left ( {\bf x} ,t \right )=\delta \left (t- {\bf x}  \right )+K\left ( {\bf x} ,t \right ),
\end{equation}
with only the rightward wave in $x\geqslant 0$. The readers can refer to Fig. \ref{fig.13}.

On the other hand, the solution with only the leftward wave in ${\bf x}\geqslant 0$ can be obtained by reversing $t$ in Eq. (\ref{eq:A22})
\begin{equation}
	f\left ( {\bf x} ,-t \right )=\delta \left (t+ {\bf x}  \right )+K\left ( {\bf x} , -t \right ).
	\label{eq:A27}
\end{equation}
Then,
\begin{equation}
	f\left ( 0, t \right )=\delta t, 
\end{equation}
and
\begin{equation}
	f_{t}\left ( 0, t \right )-f_{x} \left ( 0, t \right )=0.
	\label{eq:A29}
\end{equation}
Thus,  we have
\begin{equation}
	\begin{array}{ccl}
		u\left ( {\bf x} , t \right )&=&\int f\left ( {\bf x} , \tau \right ) u\left ( 0, t-\tau  \right ) d\tau \\
		&=& u\left ( 0, t+{\bf x}  \right )+\int_{-{\bf x} }^{\zeta }K\left ( {\bf x} , -\tau  \right ) u\left ( 0, t-\tau  \right ) d\tau \\
		&=& u\left ( 0, t+{\bf x}  \right )+\int_{-{\bf x} }^{\zeta }K\left ( {\bf x} , \tau  \right ) u\left ( 0, t+\tau  \right ) d\tau .
	\end{array}
	\label{eq:A30}
\end{equation}

\begin{figure}
	\centering
	\includegraphics[height=6cm,width=8cm]{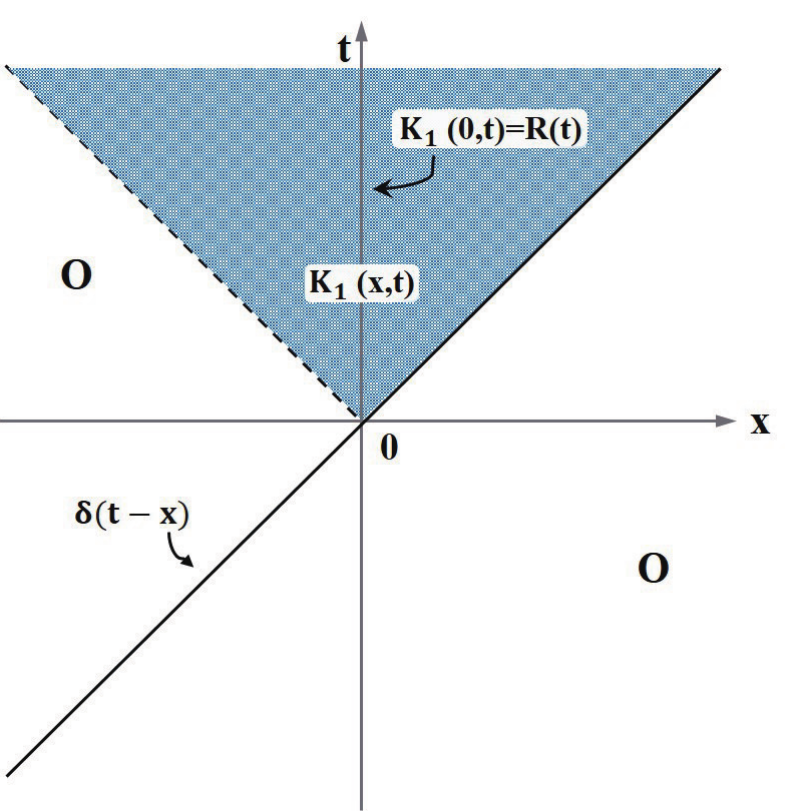}
	\caption{\label{fig.14}The causal solutions $f\left ( {\bf x} ,t \right )=\delta \left (t- {\bf x}  \right )+K\left ( {\bf x} ,t \right )$.}
	
\end{figure}

Now we consider the causal solution 
\begin{equation}
	f_{1} \left ( {\bf x} ,t \right )=\delta \left (t- {\bf x}  \right )+K_{1} \left ( {\bf x} ,t \right ).
	\label{eq:A31}
\end{equation}
The solution is shown in Fig. \ref{fig.14}. Combining Eq.~(\ref{eq:A23}) and Eq.~(\ref{eq:A27}),  we have
\begin{equation}
	f_{2} \left ( {\bf x} , t \right )\equiv K_{1}\left ( {\bf x} , t \right )- K\left ( {\bf x} , t \right ),
\end{equation}
which also satisfies Eq. (\ref{eq:A14}). Setting in Eq. (\ref{eq:A19}),  we have 
\begin{equation}
	f_{2}\left ( {\bf x} , t \right )=f_{2}\left ( 0, t+{\bf x} \right )+\int_{-x}^{x} K\left ( {\bf x} , \tau  \right )f_{2}\left ( 0,t+{\bf x}  \right )d\tau .
\end{equation}
From Eq. (\ref{eq:A16}), (\ref{eq:A17}), (\ref{eq:A21}) and (\ref{eq:A22}), we have
\begin{equation}
	f_{2}\left ( {\bf x} ,t \right )=-K\left ( {\bf x} ,t \right ),
\end{equation}
for $\left | t \right | \leqslant {\bf x}$ and 
\begin{equation}
	f_{2}\left ( 0,t \right )=K_{1}\left ( 0,t \right )=R\left ( t \right ).
\end{equation}
Substituting Eq. (\ref{eq:A30}) and (\ref{eq:A31}) into Eq. (\ref{eq:A29}), we obtain the Gel'fand-Levitan-Marchenko equation
\begin{equation}
	K\left ( {\bf x} ,t \right )+R\left ( t+{\bf x}  \right )+ \int_{-{\bf x} }^{{\bf x} } K\left ( {\bf x} ,\tau  \right ) R\left ( t+\tau  \right ) d\tau =0,
\end{equation}
where $R\left ( t \right )$ is the reflection series and $K\left ( {\bf x} ,t \right )$ is the focusing function.

The Gel'fand-Levitan Marchenko equation can be written in symbolic notation 
\begin{equation}
	K+R+\int_{W}RK=0
\end{equation}
with a time integral $\int_{W}$ and the recorded data. The fundamental  solutions describe the following scattering process. There is a source signal (incident pulse $\delta \left (t- {\bf x}  \right )$ ) which is incident at large negative time propagating along the ${\bf x}$ direction. The incident field then collides with the target (scattering potential) in the neighborhood of the origin and scatters in an outgoing wave plus the incident pulse propagating in the forward direction. Then the field is measured meaning that  we can get the scattering data which is related to the scattering objects. 

A common formulation of the wave equation can be written as 
\begin{equation}
	\left(\frac{\partial^{2}}{\partial z^{2}}-\frac{\partial^{2}}{c^{2}(z) \partial t^{2}}\right) p(z, t)=-s(t).
\end{equation}
The frequency domain wavefield is related as 
\begin{equation}
	\begin{array}{l}p(\omega, z)=\int d t \mathrm{e}^{i \omega(t-z / c)} p(t, 0) \\p(t, z)=\frac{1}{2 \pi} \int d \omega \mathrm{e}^{-i \omega(t-z / c)} p(\omega, 0)\end{array}.
\end{equation}
The continuity of waves  can be written as
\begin{equation}
	\begin{aligned}p\left(\omega, z_{-}\right) & =p\left(\omega, z_{+}\right) \\\frac{\partial p}{\partial z}\left(\omega, z_{-}\right) & =\frac{\partial p}{\partial z}\left(\omega, z_{+}\right)\end{aligned}.
\end{equation}
The time boundary condition can be written as
\begin{equation}
	\begin{aligned}p\left(t_{-}\right) & =p\left(t_{+}\right) \\\frac{\partial p}{\partial t}\left(t_{-}\right) & =\frac{\partial p}{\partial t}\left(t_{+}\right)\end{aligned}.
\end{equation}
With the time frequency transform, we derive the following equations
\begin{equation}
	\frac{\partial p(t(z))}{\partial t}=\frac{1}{2 \pi} \int d \omega(-i \omega) \mathrm{e}^{-i(\omega t-z / c)} p(\omega, 0).
\end{equation}
It follows that
\begin{equation}
	\frac{\partial p(z(t)), \omega)}{\partial t}=-i \omega p(z, \omega).
\end{equation}
Then we have
\begin{equation}
	\frac{\partial p(z, \omega)}{c(z) \partial t}=-\frac{i \omega}{c(z)} p(z, \omega)=-\frac{\partial p(z, \omega)}{\partial z},
\end{equation}

\begin{equation}
	\frac{\partial^{2} p(z, \omega)}{c^{2}(z) \partial t^{2}}=-\frac{\omega^{2}}{c^{2}(z)} p(z, \omega)=\frac{\partial^{2} p(z, \omega)}{\partial z^{2}}.
\end{equation}
We obtain the first order differential equation system
\begin{equation}
	\frac{d}{d z}\left[\begin{array}{l}p \\p_{z}\end{array}\right]=\left[\begin{array}{cc}0 & 1 \\-\frac{\omega^{2}}{c^{2}(z)} & 0\end{array}\right]\left[\begin{array}{l}p \\p_{z}\end{array}\right].
\end{equation}
The dual solution can be defined as
\begin{equation}
	\delta(t-\varsigma)+K(\varsigma, t)
\end{equation}
It follows that
\begin{equation}
	u(\varsigma, t)=\delta(t-\varsigma)+K(\varsigma, t)+R(t+\tau)+\int K(\varsigma, \tau) R(\varsigma+\tau) d \tau,
\end{equation}
Then we obtain
\begin{equation}
	0=K(\varsigma, t)+R(t+\tau)+\int_{-\varsigma}^{\varsigma} K(\varsigma, \tau) R(\varsigma+\tau) d \tau,
\end{equation}
and
\begin{equation}
	\begin{array}{l}p(\omega, x, z)=\int d t \mathrm{e}^{i \omega(t-x \sin \theta / c-z \cos \theta / c)} p(t, x, 0) \\p(t, x, z)=\frac{1}{2 \pi} \int d \omega \mathrm{e}^{-i \omega(t-x \sin \theta / c-z \cos \theta / c)} p(\omega, x, 0)\end{array}.
\end{equation}

\bibliography{aps_Timeinterface}

\end{document}